# Soft Null Hypotheses: A Case Study of Image Enhancement Detection in Brain Lesions

Haochang Shou[a], Russell T. Shinohara[b,c], Han Liu[d], Daniel S. Reich[c]
and Ciprian M. Crainiceanu[a] [*]

June 24, 2013

## Abstract

This work is motivated by a study of a population of multiple sclerosis (MS) patients using dynamic contrast-enhanced magnetic resonance imaging (DCE-MRI) to identify active brain lesions. At each visit, a contrast agent is administered intravenously to a subject and a series of images is acquired to reveal the location and activity of MS lesions within the brain. Our goal is to identify and quantify lesion enhancement location at the subject level and lesion enhancement patterns at the population level. With this example, we aim to address the difficult problem of transforming a qualitative scientific null hypothesis, such as "this voxel does not enhance", to a well-defined and numerically testable null hypothesis based on existing data. We call the procedure "soft null hypothesis" testing as opposed to the standard "hard null hypothesis" testing. This problem is fundamentally different from: 1) testing when a quantitative null hypothesis is given; 2) clustering using a mixture distribution; or 3) identifying a reasonable threshold with a parametric null assumption. We analyze a total of 20 subjects scanned at 63 visits (∼30Gb), the largest population of such clinical brain images.

[a]Department of Biostatistics, Johns Hopkins University
[b]Department of Biostatistics and Epidemiology, Perelman School of Medicine, University of Pennsylvania
[c]National Institute of Neurological Disorders and Stroke, National Institutes of Health
[d]Department of Operations Research and Financial Engineering, Princeton University
[*]The authors thank Bibiana Bielekova, Colin Shea, and the Neuroimmunology Branch clinical group for collecting and processing the data for this study. The research of Shou, Shinohara and Crainiceanu was supported by Award Number R01NS060910 from the National Institute Of Neurological Disorders And Stroke. The content is solely the responsibility of the authors and does not necessarily represent the official views of the funding agencies.

# 1 Introduction

Multiple sclerosis (MS) is a chronic immune-mediated disease of the central nervous system that results in severe disability and mortality. Localized inflammatory lesions are observed in the brains of MS patients. The disease is classified into several clinical stages, including relapsing-remitting MS (RRMS), secondary progressive MS (SPMS) and primary progressive MS (PPMS) (Fox *et al.*, 2006; Lublin & Reingold, 1996). Approximately 85% of patients initially present with RRMS involving acute attacks. These patients usually recover spontaneously within a few weeks. Following RRMS, some patients enter SPMS, in which tissue damage accumulates and disability progresses over time. In contrast, PPMS patients do not spontaneously recover but gradually worsen after disease onset.

The development of magnetic resonance imaging (MRI) has had great impact on the clinical management of MS. Characteristics of MS lesions such as volume, number, and location are crucial. In particular, dynamic contrast-enhanced MRI (DCE-MRI) is used to identify active white matter lesions through detection of associated abnormalities in the blood-brain barrier. Such abnormalities are common in patients with RRMS, but rare in subjects with SPMS or PPMS (Capra *et al.*, 1992). During each visit, a patient's brain is scanned to first acquire a few baseline MR images. After a contrast agent, gadolinium chelate, is administered intravenously, a sequence of post-injection images are obtained to show which areas of the brain have altered magnetic properties due to the presence of the contrast agent, observed as increased intensity in these regions. Post-injection images and pre-injection images are compared to identify unusual "enhancement" patterns, suggesting localized disruption of the blood-brain barrier and thus lesion activity.

As an illustration, the top and middle rows of Figure 1 show sagittal slices of $T_1$-weighted MRIs from two RRMS patients. These images are obtained by measuring the energy emitted by protons to return to their original state after being excited by a radiofrequency (RF) pulse. The RF signal decays with an exponential curve characterized by a parameter $T_1$; tissues with different paramagnetic properties appear differentiated in the image because the signal decay rate varies by tissue type. Pre-injection images are in the left column, while the middle and right columns correspond to post-injection images. The bottom row shows the same slice for subject 2 at a subsequent visit to provide information about disease progression. The estimated lesions shown in black contours are obtained from an automatic lesion segmentation algorithm (Shiee *et al.*, 2010). From pre- to post-injection images, the intensities of some lesion voxels transition from dark gray to white. Indeed, the intensity trajectories of voxels throughout the brain have various spatiotemporal patterns as shown in Figure 2. The various intensity trajectories reflect the varying characteristics of different tissues. Region-specific patterns are clear: intensities of non-enhancing lesions and normal appearing white matter (NAWM) stay roughly flat over time, while the blood vessels have a big jump immediately after the gadolinium injection and then decay quickly. Voxels in enhancing lesions increase in intensity slowly and remain higher than baseline at the end of the visit. Even for lesion voxels, the time series can vary dramatically, with newer lesions being more likely to enhance (Gaitán *et al.*, 2011).

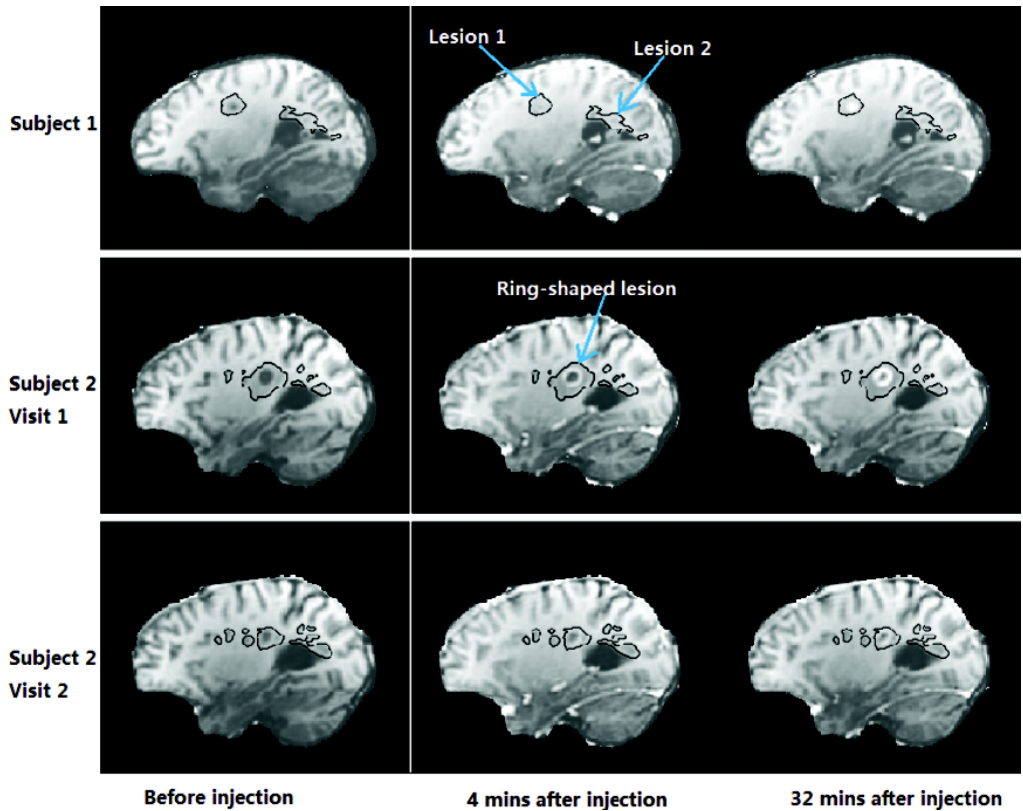

**Figure 1:** *Sagittal slices from a DCE-MRI for two subjects acquired at 3 time points before and after contrast agent injection within one visit. The images were acquired at 7 minutes before injection, and 4 and 32 minutes after injection. The black contours depict estimated lesions using Lesion-TOADS. First row: subject 1; second and third rows: subject 2 at two consecutive visits. One lesion of subject 1 and the ring-shaped lesion of subject 2 gradually light up over time. However, the ring-shaped lesion enhances less at the follow-up visit.*

Our scientific interests focus on identifying newly developed lesion voxels based on their intensity curves, and quantifying their enhancement patterns. The current practice of visual inspection by trained radiologists is effective for targeting enhancing lesion as a whole, but remains qualitative and prone to error, especially for smaller lesions and more subtle enhancement patterns. Quantitative methods for describing the temporal structure and identifying enhancing lesion voxels are important for clinical practice and for advancing our understanding of MS pathophysiology.

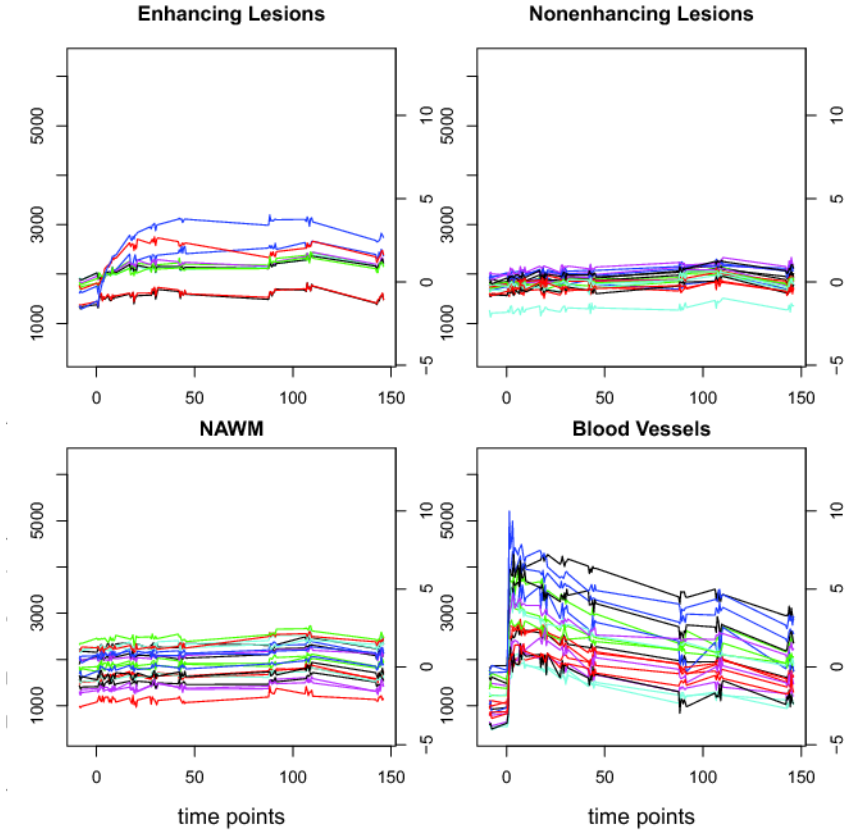

**Figure 2:** *Time series of voxel intensities in different regions of the brain from one visit. Voxels are randomly selected from the groups classified by the segmentation algorithm. Blood vessel voxels are selected based on visual inspection of images. Left axis: raw intensities; right axis: normalized intensities; x-axis: time in minutes.*

## 1.1 Experimental Design and Data Description

At each study visit, we obtain a series of $T_1$-weighted images before, during, and after injection of the contrast agent. All images are registered to the Montreal Neurological Institute (MNI) standard space. We then use a skull-stripping procedure (Carass, 2007) to remove voxels that are outside the brain. The Lesion-TOADS algorithm (Shiee *et al.*, 2010) automatically segments brain tissues including cerebrospinal fluid, gray matter, white-matter (WM) lesions and normal-appearing white matter (NAWM).

In this study, we consider 63 DCE-MRI datasets from 20 subjects. Among these patients, 4 were diagnosed with PPMS and the remaining had RRMS. Each dataset was recorded at a single study visit and many subjects were observed over multiple visits. During each visit a different number (11 to 19) of $T_1$-weighted images were recorded up to 74 minutes after the injection. One male subject (subject 1) was observed over 155 mins after injection at one single visit, and 67 images were acquired to assess long-term enhancement behaviors. The $T_1$-weighted images from each visit are stored in a 4-dimensional array. The first 3 coordinates represent the right-left, posterior-anterior, and inferior-superior axes and are of dimension $182 \times 218 \times 182$ for a total of more than 7 million entries. The 4th dimension is the time at which the images are acquired during the visit and varies by scan. The largest DCE-MRI for a subject at one visit in our dataset had almost 500 million data points, while an average DCE-MRI had roughly 100 million entries. The time points are calibrated so that the contrast agent injection occurs at time 0. The segmentation information is stored as 3-dimensional binary arrays indicating which voxels correspond to which tissue type. To avoid any misunderstanding, it is worth mentioning that DCE-MRI is a completely different modality from functional Magnetic Resonance Imaging (fMRI). The scientific problems addressed using DCE-MRI are fundamentally different from those of fMRI. To the best of our knowledge, there is currently no publication in the statistical literature dedicated to one full-brain DCE-MRI, not to mention populations of such images.

## 1.2 Previous Work

The state-of-the-art for analyzing DCE-MRI is to apply pharmacokinetic (PK) (Davidian & Giltinan, 1995) models at every voxel (Tofts *et al.*, 1999; Yankeelov *et al.*, 2005; Li *et al.*, 2005). Through a set of PK parameters, the model connects the observed intensity time course with dynamic concentration of contrast agent and tissue characteristics (Tofts *et al.*, 1999). The extracted tissue characteristics are then used to identify abnormal behavior in lesions. The appeal of PK models for DCE-MRI is that they are intuitive, simple, well-understood, and derived from well-defined physical models. However, PK models fail on more than 80% of the brain voxels in our example. Perversely, they fail exactly where quantification is most important: in areas of moderate and small enhancement in and around lesions. A quick inspection of the data displayed in Figure 2 should explain why this is the case: models that work in the middle of the lesion may be completely inadequate in the normally appearing white matter, blood vessels, or areas immediately surrounding the lesion. When the focus is on estimating the extent and type of lesion enhancement, this is a fatal flaw. Moreover, PK models make assumptions that are hard to defend in practice: in particular, that a fixed number of compartments is enough for every voxel irrespective to their tissue types. Our experience is that the standard single-compartment PK model works well in practice *after lesions are identified* and only for those *voxels with pronounced enhancement*. Here we are concerned with automatic identification of voxels that enhance and characterization of enhancing behavior along its continuum: no enhancement, small, moderate and strong enhancement.

Another problem is that standard approaches focus on analyzing one subject at a single visit. Here, we are interested in characterizing populations of such visits and characterizing the population-level structure of enhancement patterns. Shinohara *et al.* (2011) proposed a different approach using functional principal component analysis (FPCA) to extract population-level features of lesion enhancement. Using data from 10 subjects observed cross-sectionally, their work suggested that lesion enhancement patterns were captured by two principal components. In our paper, we first perform a confirmatory analysis of the findings in Shinohara *et al.* (2011) using a dataset that is roughly 6 times larger. We then focus on our main goal: designing a null hypothesis and a test that are consistent with the visual inspection (qualitative assessment) in finding enhancing lesion voxels. Once this is done in several subjects we investigate the generalizability of our approaches to other subjects.

## 1.3 Soft Null Procedure

The classical hypothesis testing aims to assess the congruence of the observed data with a formal and testable hypothesis of interest, which we refer to as a "hard" null hypothesis. However in our example, one qualitative scientific hypothesis may be interpreted in varying ways which correspond to a set of rigorously defined null hypotheses. Defining and choosing a hard null hypothesis is difficult and the result may depend heavily on this choice.

We propose using soft null hypotheses as a framework for assisting the transition from a scientific notion to a well-defined hypothesis that is appropriate for testing. In this procedure, data play an important role in generating, refining and selecting this hard null from a sequence of candidate "soft" null hypotheses. Although current investigators may have been using information from data, consciously or not, to build up hard null hypothesis, the framework that we describe formalizes this process and highlights its importance. The soft null procedure consists of the following steps: 1) quantify the qualitative hypothesis, which provides a likely set of candidate null hypotheses; 2) test the candidate hypotheses on a training dataset; 3) based on the testing results, reevaluate the the candidate hypotheses and generate more refined hypotheses; 4) identify a few hypotheses as suitable hard null hypotheses; 5) test the hard null hypotheses on test data in the classical inference paradigm. Using the application of lesion enhancement detection to illustrate the details of soft null procedure, we aim to add some structure and statistical rigor to "hypothesis generation", which is an ill-defined process usually left to scientist who are not assisted by statisticians.

Specifically, our objective is to refine quantitative approaches to testing enhancement at the voxel level. Figure 1 indicates that many parts of the brain, such as the blood vessels and meninges also enhance. Moreover, lesion locations are not accurately estimated. Thus, defining the null hypothesis of normal intensity change in lesions is difficult. The problem with identifying a mathematically rigorous null hypothesis stems from how we naturally treat information. For example, the word "enhancement" qualitatively describes a physical property of brain tissue as it characterizes a transition from shades of black to shades of white in images, such as in Figure 1. However, precise formulation of "enhancement" needs to undergo a process of generating and refining the definition of possible null hypotheses. In Section 4, we investigate five biologically meaningful working null hypotheses and estimate the associated null distributions. Three of the definitions are based on visit-specific data and two are based on population-level analyses. These definitions of the null hypothesis corresponding to "no enhancement" have different interpretations and incorporate various levels of additional information.

## 2 Validation of Established Techniques

Two problems must be addressed before testing for enhancement. First, raw MR intensities are not normalized within or across subjects, making population-level analyses and generalizations difficult. Second, data are very large and require dimensionality reduction. For the first problem, Shinohara *et al.*, 2011 proposed normalization with respect to NAWM intensities measured before gadolinium injection. The procedure is simple and intuitive, only requiring data within the same visit. For the second issue, Shinohara *et al.*, 2011 used PCA of all voxel-level time series across 10 subjects. They conclude that: 1) the first four principal components (PCs) explain more than 99.8% of total variability; 2) two principal components capture lesion-enhancing behavior; and 3) analyzing the scores of these two principal components enables quantifying time series behavior across regions of interest. We deploy their approach on a much larger dataset (63 versus 10 scans), and investigate the reproducibility of conclusions made in this previous work.

In particular, the images are normalized as $Y_i(t,v) = \frac{Y_i^O(t,v) - \mu_{i,0}}{\sqrt{V_{i,0}}}$, where $Y_i^O(t,v)$ is the raw intensity of voxel $v$ at within-visit time $t$ in visit $i$, $Y_i(t,v)$ is the associated normalized intensity, and $\mu_{i,0}$ and $V_{i,0}$ are the sample mean and variance for the pre-injection intensities of all NAWM voxels. In addition, the inconsistent imaging acquiring time across visits enables us to characterize population-level features in the data throughout the scanning period, particularly during the first hour. Given the sparsity of the sampling grids across subjects, we smooth the time series for each voxel at every visit using linear interpolation as suggested by Shinohara *et al.* (2011) and obtain intensity measurements on an equally-spaced time grid. The details are omitted here.

After normalizing the intensity and interpolating the time series, we conduct population-level PCA to capture the temporal patterns of voxel intensities in different brain regions. The first nine population-level principal components (PLPC) (Figure 3) resemble those obtained in smaller sample size. As the earlier data is a subset of our dataset, we conduct the same analysis excluding those in Shinohara *et al.*, 2011 and obtain similar results (omitted). Smoothing the covariance operator may be done before eigenvalue decomposition. However, this was unnecessary due to the large number of voxels and dense sampling across subjects.

The first four PLPCs explain more than 99.5% of the total variance. PC 1 is flat over time and accounts for heterogeneity of baseline intensities across different brain regions. The second PC describes an instantaneous increase in voxel intensity after injection, reflecting contrast dynamics in blood vessels (see, for comparison, the bottom right panel in Figure 2). The third PC displays a slow increase over time and corresponds to the gradual brightening of some voxels within enhancing lesions. The fourth component shows a fast increase followed by a fast decline in voxel intensity, peaking at approximately 30 minutes after injection. After the PLPCs have been estimated, the PC scores for voxel $v$ in dataset $i$ are obtained as $\xi_{ik}(v) = \langle Y_i(t,v), \phi_k(t) \rangle \approx \sum_{i=1}^{T} Y_i(t_i, v)\phi_k(t_i)$, where $Y_i(t,v)$ is normalized intensity at time $t$ for voxel $v$ of subject $i$. The scores are then mapped back to the brain (Figure 4) where spatial correlation is reconstructed. Figure 4 displays maps of the first 4 PC scores for visit 1 of subject 2 (middle panel of Figure 1) on four different slices, confirming interpretation of the PLPCs. We conclude from Figure 4 that voxels which gradually enhance over time have high projections on the $3^{\text{rd}}$ and $4^{\text{th}}$ PCs.

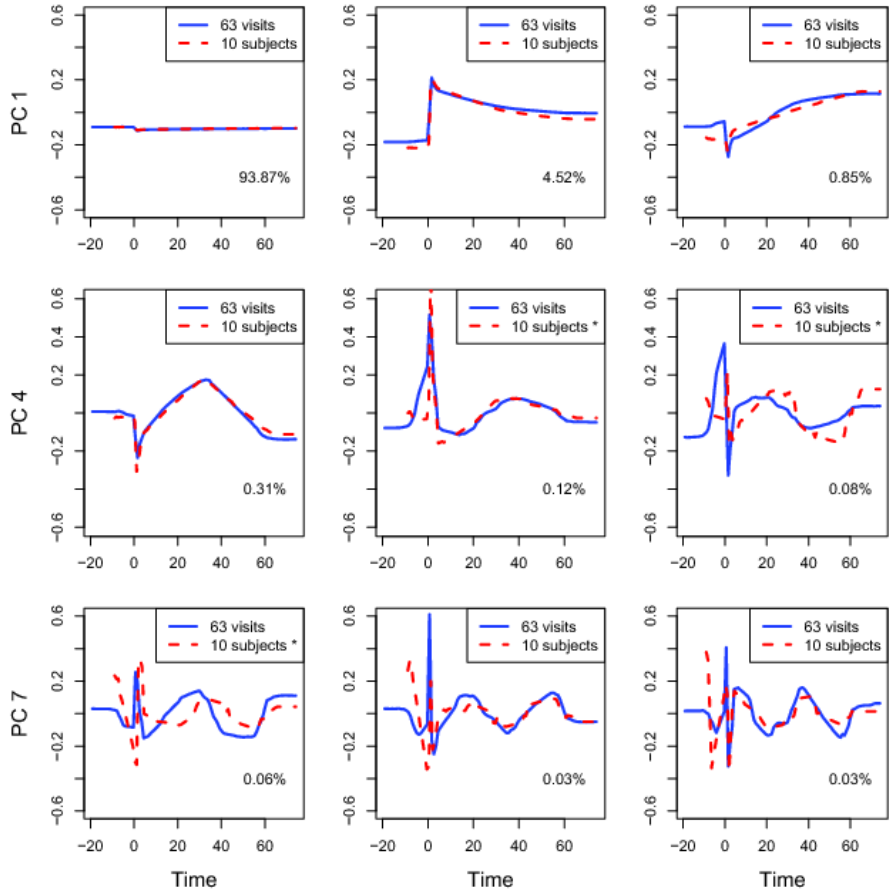

**Figure 3:** *The first nine principal components from the population-level analysis of the 63 visits are shown in blue lines. The numbers in percentage at the bottom right corner are the proportions of variance explained by each component. The dashed red lines show the first nine PLPCs obtained by the 10-subject analysis from Shinohara et al. , 2011. The first four curves are almost identical. Since the PC components are invariant with respect to sign, the $5^{th}$ to the $7^{th}$ components from the 10-subject results were flipped to match with the PLPCs from the larger analysis.*

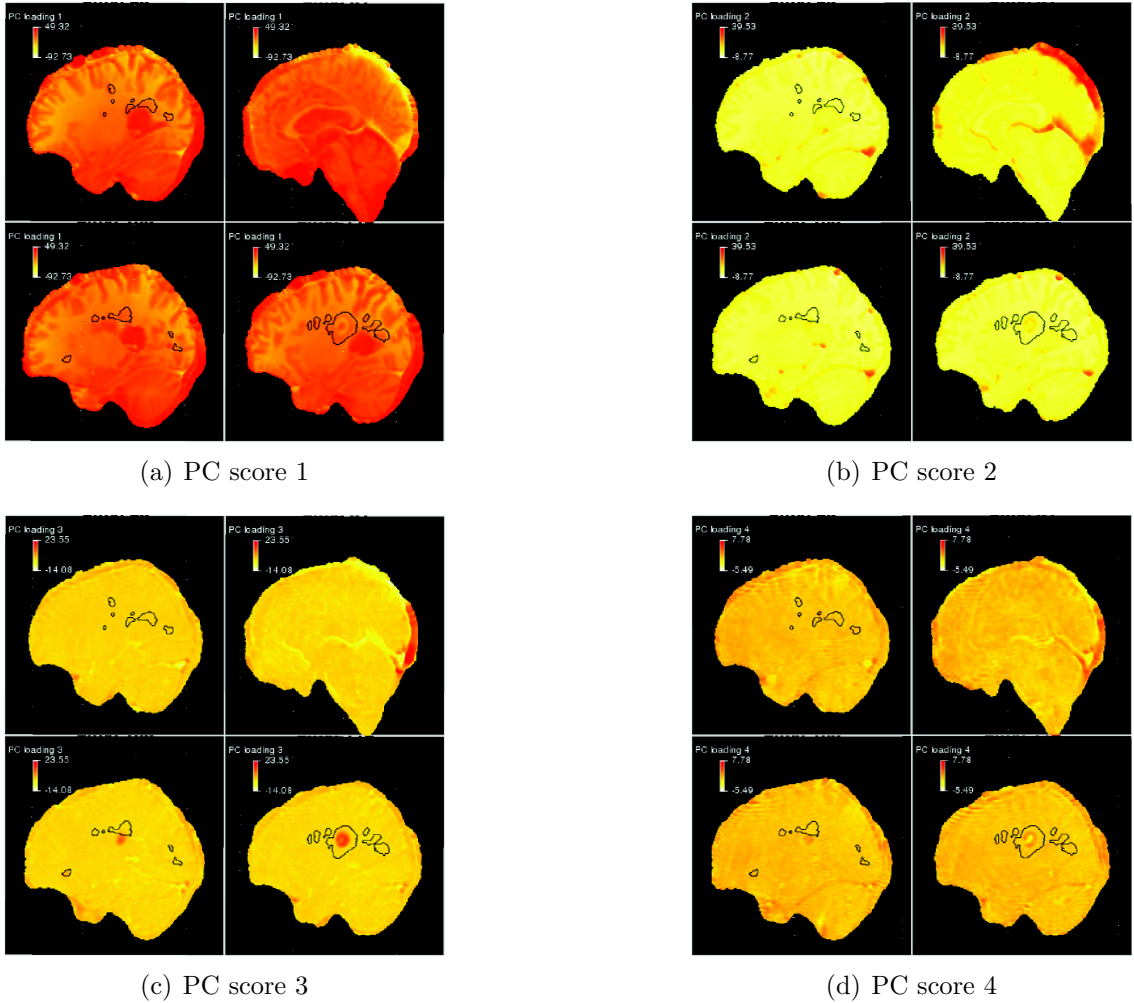

**Figure 4:** *PC score images in the brain of the second subject in four sagittal slices. Voxels with high $2^{\text{nd}}$ PC scores are mostly in blood vessels. Some lesions (within the black contours), however, also have high PC2 scores. The ring-shaped lesion contains voxels with high PC3 and PC4 scores, but some of the voxels near the back of the brain have have similar scores.*

To quantify enhancement, we restrict our analysis to the bivariate PC score space $\mathcal{S}_i = \left[ \{\xi_{i3}(v), \xi_{i4}(v)\}, \, v = 1, 2, ..., V_i \right]$. The left panel of Figure 5 displays scatter plots of these pairs of PC scores for 5 subjects (each subject shown in a different color). These plots contain a dense central area with long right tails shooting towards the top-right corner of the plot. Thus, in this space, the problem becomes finding a null hypothesis and a decision boundary that discriminates between lesion enhancement voxels and other voxels. We argue that this problem requires care-ful hypothesis development, testing on data, followed by refinement of both hypothesis and test statistics. As no statistical textbooks discusses this data-based process that applied Statisticians have been using forever we contend that this is a new theoretical concept.

## 3 Clustering with Mixture Distributions

In the search for the right testing framework we start with a comfortable approach: using mixture distributions and hoping for the best. Consider the case when we fit a sequence of mixtures of bivariate normal distributions for one of the subjects corresponding to an increasing number of mixture components. The right panel of Figure 5 displays results from fitting a mixture of 2 bivariate normal distributions. The solid-black ellipsoids indicate the contours of one standard deviation away from the mean for the two normal distributions. The number of voxels that are estimated to enhance (the blue dots) is very large. Plotting these voxels back on the brain template would provide a case of the "blue brain", one that is estimated to enhance almost everywhere. We have also conducted clustering with three to six components. Given the huge number of voxels in the brain, a larger number of components is always preferred using standard criteria. However, even with an increased number of clusters, identifying the cluster of enhancing lesion voxels remains elusive (See supplementary materials for details). An additional practical difficulty is that as we increase the number of clusters, cluster assignments of voxels change dramatically.

In this context using FDR instead of Bonferroni correction would be an even worse approach because it would simply allow for more voxels to be estimated as enhancing. This is a fundamentally different problem from the studies discussed in Efron (2004, 2007). In their case the problem is to allow for a more liberal $\alpha$ level in situations where the signal is very weak. Here we are dealing with millions of voxels that do not enhance at all, most of which are concentrated around the origin, mixed with hundreds of thousands of lesion- and non-lesion voxels whose patterns of enhancement span the continuum from very subtle to ultra-obvious. From a statistical perspective we are thus confronted with a subject-specific complex mixture of an unknown number of distributions with unknown characteristics. To our knowledge, no such problem has been posed and standard statistical approaches do not provide a solution. Without a rigorous definition of "non-enhancement" that can pass visual inspection of data and results, we quickly run into circular logic. In the next section, we will take the testing approach and consider a sequence of working null hypotheses that quantify the scientific hypothesis of detecting enhancing lesion voxels.

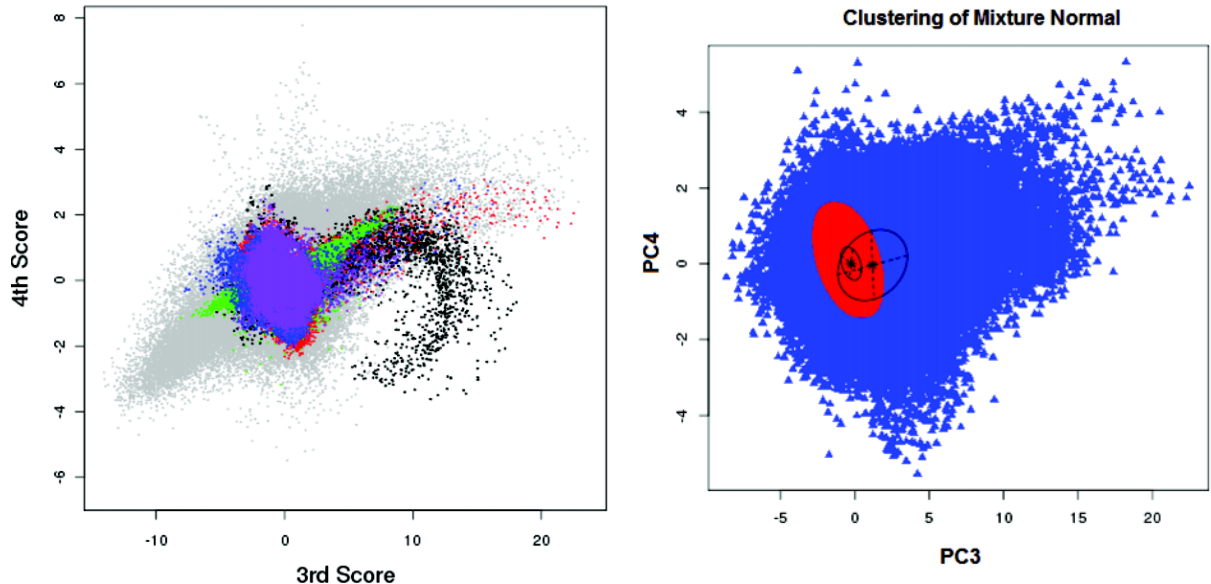

**Figure 5:** *On the left panel shows the scatterplot of $3^{\text{rd}}$ and $4^{\text{th}}$ PC scores from 5 subjects and 1 visit per subject. Voxels in WM masks are colored by visit, and other voxels are shown in gray. Most of the dots are concentrated around the origin, meaning that the majority of voxels have relatively small PC 3 and 4 scores; While a small group of voxels have extreme PC scores, shown in the plot as the arms shooting towards the upper-right corner. Also note that data from 5 subjects overlap with each other quite well, which suggests that the normalization procedure corrects most of the heterogeneity across different data. On the right we have the clustering results using a mixture of two bivariate normals for the $3^{\text{rd}}$ and $4^{\text{th}}$ PC scores from one dataset. Two ellipsoids: the two normal distributions (mean plus one standard deviation). Red dots: estimated non-enhancing voxels. Blue dots: estimated enhancing voxels. Bonferonni correction was used.*

# 4 Testing for Enhancement

Through statistical hypothesis testing, we aim to determine voxels that enhance abnormally, a behavior associated with active inflammation in newly formed lesions. However, we identified two main difficulties: lack of a precise definition of non-enhancement and lack of methods for estimating the associated null distribution. Generally considered a purely scientific endeavor, the first problem is more difficult and conveniently dismissed.

In this section, we navigate through several definitions of non-enhancement and propose simple methods for estimating the corresponding null distributions. We first present a spectrum of null hypotheses specific to each dataset, starting with the narrowest definition of non-enhancement and expanding towards more liberal definitions. Each null hypothesis corresponds to our scientific goal of enhancement detection, but each leads to fundamentally different interpretations. We also consider two null hypotheses defined across subjects that incorporate population heterogeneity.

## 4.1 Null distribution based on a single visit

We propose several approaches that define null distributions estimable from the data measured at a single study visit. Such a visit-specific null distribution avoids issues caused by inappropriate normalization across subjects and visits.

### 4.1.1 $H_{0,1}$ Counterfactual Null

We first consider the following null:

$H_{0,1}$: voxel-specific time series dynamics is the same before and after contrast injection

Let $Y_i(t,v)$ be the intensity of voxel $v$ at time $t$ at visit $i$, $\mu_{i,<0}(v)$ and $\sigma^2_{i,<0}(v)$ be the voxel-specific mean and variance of intensities measured before time 0 (injection). For voxels where $H_{0,1}$ holds, their intensity should have the same distribution before and after injection. In particular, these voxels will have the property $\mathcal{A}_{1,i} := \{v : \mathbb{E}[Y_i(t,v)] = \mu_{i,<0}(v), Var[Y_i(t,v)] = \sigma^2_{i,<0}(v) \text{ for all } t\}$. We call $H_{0,1}$ the counterfactual null hypothesis because it defines non-enhancement based on the intensity time series of voxels when gadolinium injection is absent (not observed in our study).

We now focus on estimating the bivariate distribution of the 3$^{\text{rd}}$ and 4$^{\text{th}}$ PC scores for non-enhancing temporal patterns under $H_{0,1}$. We assign each voxel a non-enhancing 'match' by simulating a counterfactual time series based on pre-injection observations:

$$Y_i^{\text{null}}(t,v) = \bar{\mu}_{i,<0}(v) + s_i(t,v), \tag{1}$$

where $\bar{\mu}_{i,<0}(v) = \frac{1}{\sum_t I(t<0)} \sum_{t<0} Y_i(t,v)$ is the averaged pre-injection intensity that estimates $\mu_{i,<0}(v)$, $I(t<0)$ is the indicator function for $t<0$, and $s_i(t,v)$ is randomly sampled from the collection of residuals of all the voxels before time 0; namely, $s_i(t,v) \in \{r_{i,<0}(u) = Y_i(s,u) - \bar{\mu}_{i,<0}(u) : s < 0, u = 1, 2, \cdots, V_i\}$. Figure 6(a) displays 7 sample time series from the original dataset together with their simulated counterfactual 'matches'. This procedure assumes that the errors deviating from the pre-injection mean are white noise and follow a common distribution across all voxels. Figure 6(b) supports this assumption as the standard deviation of the pre-injection intensities for each voxel is roughly independent of its mean and does not vary across tissue types.

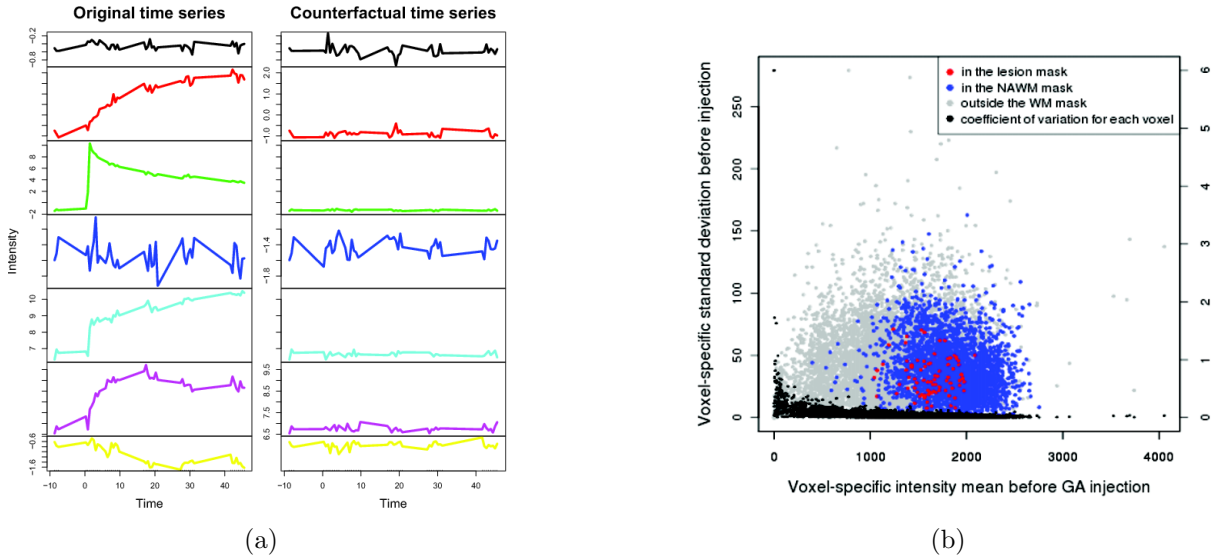

**Figure 6:** *(a) Time series of seven voxels from a single subject and their counterfactual curves generated under $H_{0,1}$. (b) Standard deviations of $\{r_{i,<0}(v)\}$ (y-axis) versus $\bar{\mu}_{i,<0}(v)$ (x-axis). Each point corresponds to one voxel in the brain. The left axis is in the raw intensity scale and the right axis is in the normalized scale. The color indicates tissue types according to the segmentation information available. Black dots are the coefficients of variation calculated for each voxel. The standard deviation does not seem to depend on the tissue type or on the mean intensity.*

After simulating the non-enhancing counterpart for all voxels $\widetilde{\mathbf{Y}}_i := \{Y_i^{\text{null}}(t',v)\}_{V_i \times T}$, we project them onto the $k^{\text{th}}$ ($k = 3, 4$) component space and obtain the 'nullified' PC score for voxel $v$ as $\xi_{ik}^{\text{null}}(v) := \langle Y_i^{\text{null}}(\cdot,v), \phi_k(\cdot) \rangle \approx \sum_t Y_i^{\text{null}}(t,v)\phi_k(t)$. We then estimate the density of the null distribution based on $\{\xi_{i3}^{\text{null}}(v), \xi_{i4}^{\text{null}}(v)\}$. For our analysis, we fit a bivariate normal distribution $N(\widehat{\mathbf{u}}, \widehat{\boldsymbol{\Sigma}})$ to these scores. The $p$-value for voxel $v$ can be interpreted as the probability of observing a sample from $N(\widehat{\mathbf{u}}, \widehat{\boldsymbol{\Sigma}})$ that is more extreme than $\{\xi_3(v), \xi_4(v)\}$. Specifically, we define:

$$p_v = \begin{cases} \Phi\big(\xi_3(v) < \xi_3, \xi_4(v) < \xi_4 | \widehat{\mathbf{u}}, \widehat{\boldsymbol{\Sigma}}\big), & \text{if } \xi_3(v) > u_3 \text{ and } \xi_4(v) > u_4; \\ \Phi\big(\xi_3(v) < \xi_3, \xi_4(v) > \xi_4 | \widehat{\mathbf{u}}, \widehat{\boldsymbol{\Sigma}}\big), & \text{if } \xi_3(v) > u_3 \text{ and } \xi_4(v) \leq u_4; \\ \Phi\big(\xi_3(v) > \xi_3, \xi_4(v) < \xi_4 | \widehat{\mathbf{u}}, \widehat{\boldsymbol{\Sigma}}\big), & \text{if } \xi_3(v) \leq u_3 \text{ and } \xi_4(v) > u_4; \\ \Phi\big(\xi_3(v) > \xi_3, \xi_4(v) > \xi_4 | \widehat{\mathbf{u}}, \widehat{\boldsymbol{\Sigma}}\big), & \text{if } \xi_3(v) \leq u_3 \text{ and } \xi_4(v) \leq u_4. \end{cases} \quad (2)$$

where $\Phi(\cdot|\widehat{\mathbf{u}}, \widehat{\boldsymbol{\Sigma}})$ is the cumulative distribution function of $N(\widehat{\mathbf{u}}, \widehat{\boldsymbol{\Sigma}})$ with $\widehat{\mathbf{u}} = (u_3, u_4)^T$. At 95% confidence level under Bonferroni correction, we reject $H_{0,1}$ for voxels whose $p_v < 0.05/V_i$, where $V_i$ is the number of voxels for subject $i$ and is typically in the millions. Given the observed strong enhancement patterns in the brain, doing a Bonferonni correction is not too conservative. On the contrary, we believe that a correct definition of the null hypothesis is the single most influential factor in concluding scientific results. We exclude voxels with negative $\xi_3$ or $\xi_4$ since we are only interested in those with high scores. $\mathcal{R}_{1,i} = \big\{v : p_v < 0.05/V_i, \xi_3(v) > 0, \xi_4(v) > 0\big\}$ is the rejection region.

Figure 7 summarizes the testing results of $H_{0,1}$ for the first subject. In 7(a), the $p$-value of each voxel is coded using the color scheme shown in the legend. The red dots are amplified for clarity. The black ellipsoid indicates the Bonferroni-corrected 95% confidence region for the null distribution under $H_{0,1}$. Figure 7(b) shows the location of the rejected (enhancing) voxels in four different slices of the brain. On slice 65, part of lesion 1 is successfully identified with extremely small $p$-values. Some voxels in the second lesion are also selected (shown in slice 72). On slice 96, a group of orange dots near the forehead indicates an enhancing lesion that was not detected by the automated lesion segmentation method, but is discovered by our method. It was later confirmed through visual inspection by an expert neuroradiologist with extensive experience with MS. In 7(d), we stratify the voxels into four groups by their $p$-values, draw 20 random samples from each group and plot their time series (shown in color corresponding to $p$-value) as well as the average intensity curve (black). Small $p$-value (i.e. significant result in the testing) is consistent with large degree of enhancement in the observed time series.

This approach works reasonably well. However, we find that $\{\xi_3^{\text{null}}, \xi_4^{\text{null}}\}$ are highly concentrated around the origin and $\widehat{\boldsymbol{\Sigma}}$ is quite small (the black ellipse in 7(a)). This results in a rejection region that is too large and leads to many false positives; see Figure 7(b). More precisely, in this example, $H_{0,1}$ is rejected at 5.7% of the whole-brain voxels. Among the rejections, 0.88% are labeled as "within-lesion" by the Lesion-TOADS algorithm, 0.83% are in the NAWM mask and the rest fall outside the white matter mask. In fact $H_{0,1}$ is rejected in many regions that are not of biological interest, including blood vessels, non-enhancing lesions and even NAWM (see, for example, slice 90 in 7(b)). We conclude that pre-injection behavior of a voxel cannot accurately approximate its post-injection behavior, even if that voxel is not in an enhancing lesion. If we constrain the candidate voxels to be within the white matter mask, many false positive voxels can be excluded; for example, comparing Figure 7(c) with 7(b) indicates that such a procedure would eliminate the voxels near the meninges or around larger veins. However, many of the voxels of the new lesion that we identified in the forehead would also be excluded, which would prevent a scientific finding. An additional problem with this type of analysis is that white matter masks from automated methods such as Lesion-TOADS are far from perfect and in many studies may not even be available. Therefore, we remain on the whole-brain analysis.

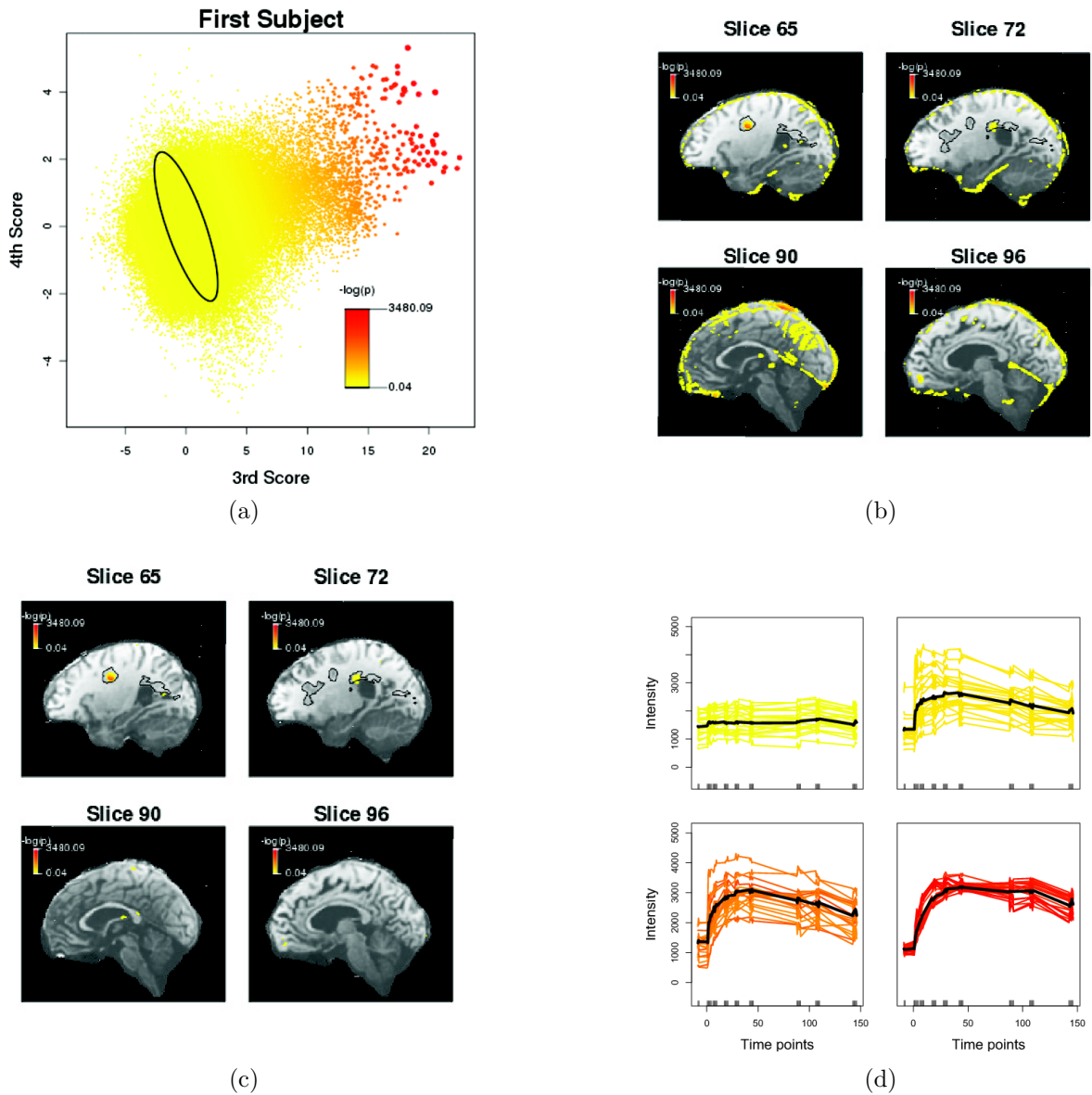

**Figure 7:** *Analysis results of data for subject 1 using the counterfactual null. (a) The p-value of each voxel is represented in color and the size of red dots are amplified for clarity. The black ellipsoid is the 95% confidence contour under the Bonferroni correction using the total number of voxels that are tested; (b) The location of rejected voxels in four different slices of the brain. Slice 65 shows that voxels in the enhancing lesion are successfully classified. Slice 72 shows that part of another lesion is also identified. There are quite a few white matter voxels that are selected in slice 90. On slice 96, a group of orange-colored voxels correspond to an enhancing lesion that is not detected by the automated lesion segmentation method, but is discovered by our method; (c) Voxels that are in the white-matter mask and have significant p-values, shown on the same four slices as in (b); (d) We stratify the voxels by their p-values into four groups and draw 20 random sample series in each group. The trajectories shown in color are the observed time series and the mean curve in black. The upper-left graph shows voxels that are not deemed enhancing.*

Although the counterfactual null is liberal for detecting enhancing lesions, it provides us an unique exploratory tool that assists in understanding the actual enhancement mechanism and quantifying enhancement pattern from a single scan when no other segmentation information is available. Indeed, the testing results from $H_{0,1}$ motivate a new set of hypotheses $H_{0,2}$ and $H_{0,3}$ which account for the intensity change in normal regions.

### 4.1.2 $H_{0,2}$ NAWM-based Null

We have seen that even voxels in NAWM may become brighter after the gadolinium injection and have positive 3rd and 4th PC scores. This is due to normal blood flow to the tissues through capillaries or small vessels. To diminish false positives, we need to account for these normal intensity changes when building null hypotheses. This is possible when a NAWM mask is available (from a segmentation algorithm) for each visit. Our null hypothesis may then be expressed as:

$H_{0,2}$: the voxel-specific time series dynamics is the same with those of NAWM voxels

As a result, for voxel $v_0$ where $H_{0,2}$ is true, $\{\xi_3(v_0), \xi_4(v_0)\}$ follows the same distribution as in $\mathcal{A}_{2,i} := \{[\xi_3(v), \xi_4(v)], v \in \text{NAWM}\}$. As the true NAWM segmentation is unknown, we estimate $\mathcal{A}_{2,i}$ by $\widehat{\mathcal{A}}_{2,i} = \{[\xi_3(v), \xi_4(v)], v \in \text{NAWM masks estimated by the Lesion-TOADS algorithm}\}$. Similar to the previous approach, we estimate the null distribution by fitting a bivariate normal to $\widehat{\mathcal{A}}_{2,i}$ and calculate $p$-values for each voxel.

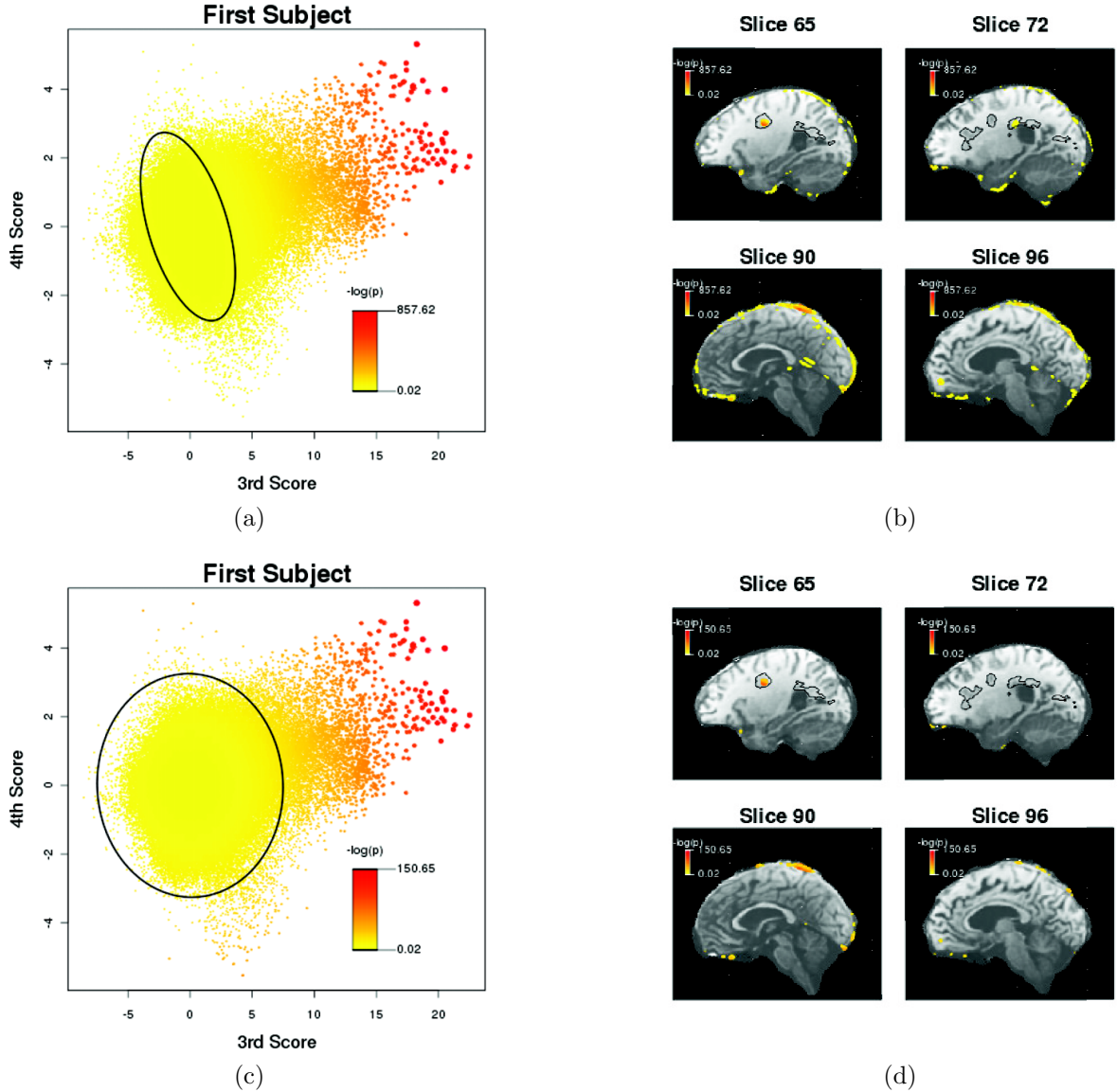

**Figure 8:** *(a-b) Analysis results of data for subject 1 using the NAWM-based null $H_{0,2}$. (a) The p-value of each voxel is represented in color and the size of red dots are amplified for clarity. The black ellipsoid is the 95% confidence contour under the Bonferroni correction using the total number of voxels that are tested. A much broader confidence region is generated than Figure 7(a). (b) The location of rejected voxels in four different slices of the brain. Note that less white matter voxels are selected on slice 72 compared to Figure 7(b). Nevertheless, the enhancing voxels in lesion mask and in the detected enhancing lesion missed by Lesion-TOADS are still selected. (c-d) Analysis results of data for subject 1 using the complementary null $H_{0,3}$. This approach is more conservative in terms of defining enhancement compared to the first two null hypotheses. Note that on slice 72 in (d), voxels in the second lesion are no longer significant.*

Compared to the counterfactual null, $H_{0,2}$ generates a much broader 95% confidence region (black contour in Figure 8(a)). For the same data from subject 1, 2.38% of the voxels in the brain are rejected, among which 1.4% are in the lesion masks and 0.19% are in the NAWM mask. The main difference between Figure 7(b) and 8(b) is on slice 90 where most of the NAWM voxels that are identified as enhancing in $H_{0,1}$ are filtered out when testing $H_{0,2}$. While selecting fewer false positive voxels, the enhancing voxels in the lesion areas are still identified and are shown on slice 65, 72 and 96 of Figure 8(b).

### 4.1.3 $H_{0,3}$ Complementary Null

$H_{0,2}$ successfully identified enhancement in lesions and eliminated most of the NAWM voxels falsely classified using the counterfactual null. However, it fails to separate voxels in the enhancing lesions from those in blood vessels and around the meninges. We conclude from Figure 5 that this is inevitable if we base our analysis only on $\{\xi_3(v), \xi_4(v)\}$ but no additional information. This is because the PC scores for voxels outside the WM mask (gray dots in Figure 5) spread as widely as lesion voxels (the upper-right tail of the colored dots). To solve this, we consider including both NAWM and voxels outside white matter in the null distribution estimation. The corresponding null hypothesis for this case is:

$H_{0,3}$: the voxel-specific time series dynamics is the same with that of non-lesion voxels

We call $H_{0,3}$ the complementary null because it is defined based on the complement set of the lesion masks where we are interested in searching for enhancement. We denote the null set of PC scores under $H_{0,3}$ by $\mathcal{A}_{3,i} := \{[\xi_{i,3}(v), \xi_{i,4}(v)], v \text{ is not in lesions}\}$. Similar to $H_{0,2}$, we estimate $\mathcal{A}_{3,i}$ by $\{[\xi_{i,3}(v), \xi_{i,4}(v)], v \notin \text{lesion masks}\}$.

The bottom row of Figure 8 shows the results from applying the test to the first subject. Compared with $H_{0,2}$, a smaller number of voxels (0.37% of the whole brain) are rejected and a larger proportion (4.03%) of the rejections are located in the lesion masks. $H_{0,3}$ is more conservative in terms of defining enhancement compared to the first two null hypotheses and generates the smallest rejection region among the three visit-specific null hypotheses. On slice 72 in 8(d), voxels in the second lesion are no longer selected as significant, but the forehead lesion is still detected. In addition, the number of false positive voxels around the edge of the cerebrum and in the vessels is greatly reduced. Note that through comparing the testing results of different null hypotheses using data from one subject and combining it with spatial information, we are able to better understand the qualitative scientific null hypothesis.

Both $H_{0,2}$ and $H_{0,3}$ rely on the segmentation masks for each dataset. A natural question would be: how much do results depend on the segmentation accuracy? Indeed, given a perfect lesion segmentation we could restrict our analysis to the lesion mask. However, all segmentation algorithms are imperfect. Instead of applying the lesion masks to filter the final testing result, we incorporate the mask in the null distribution estimation. We believe that the segmentation is correct for most of the voxels and the correctly labeled voxels form a representative sample from the population. As a result, density estimation is not severely biased by the misclassification. In fact, the testing results can conversely be used to tune the segmentation algorithms. For example, as shown in the supplementary document, testing results for subject 2 reveals that one lesion on slice 109 has an enhancing region that lies outside the original lesion mask. After visual inspection our collaborators indicated that the mask was indeed flawed and that our approach correctly identified the extent of the lesion enhancement.

As we progress from $H_{0,1}$ to $H_{0,3}$, the definition of null hypothesis is refined and various comparisons across the results can be investigated. The sequential hypotheses testing procedure unveils various degrees of dynamic changes in voxel intensities throughout the brain. Despite the consistency in detecting lesion 1 as well as the frontal lobe lesion, the associated non-rejection areas at the same $\alpha$-level become larger and the proportion of rejections in lesion masks increases in accordance with visual inspection. If interest lies in a more liberal partition of voxels which detects diffuse enhancement throughout the brain, then $H_{0,1}$ may be ideal. However, if isolating lesion voxels that have dramatic enhancement properties is the goal, $H_{0,2}$ or $H_{0,3}$ are more appropriate. The scenarios presented in this section are defined within a single DCE-MRI dataset. In most studies, however, data from multiple subjects and visits are collected; in the next section we consider using data across subjects to define non-enhancement.

## 4.2 Population-level Null Distribution

Incorporating information from multiple subjects and visits is key in classical statistical problems for evaluating variability in a population. Here, we consider formulating null hypotheses that borrow strength across datasets to estimate null distributions.

### 4.2.1 $H_{0,a}$ NAWM across Samples

We start with an approach that is similar to $H_{0,2}$. Using data from multiple subjects accounts for population heterogeneity that is not addressed by normalization. It also allows us to apply the estimated null distribution directly to future images acquired under the same protocol and avoids requiring segmentation results for the particular visit being analyzed. In particular, we define the null hypothesis in terms of the intensity behavior from voxels in the NAWM masks of all datasets other than the testing data.

> $H_{0,a}$: the voxel-specific time series dynamics is the same with with those of NAWM voxels in the population

Let $\mathcal{A}_a := \{[\xi_{j,3}(v), \xi_{j,4}(v)], I_{\text{NAWM}}(v) = 1, \ J \in I\setminus\{i\}\}$, where $J$ is the index for all datasets. To estimate the null distribution under $H_{0,a}$, we fit a normal distribution to the 3rd and 4th PC scores for all voxels which are classified as NAWM from visits other than the one to be tested. We then apply the testing procedure developed previously. Taking the first subject as an example (results shown in the upper row of Figure 9), we reject 0.47% of the voxels in the brain, among which 3.52% are in the lesion mask and 0.34% are in the NAWM mask. With a larger number of voxels that can be used for density estimation, this paradigm provides more stable results than the visit-specific cases from Section 4.1. Compared with $H_{0,2}$ (Figure 8(b)), $H_{0,a}$ classifies far fewer voxels around the meninges and vessels as "enhancing". Even though these voxels are not contained in the NAWM masks and do not contribute in estimating the null distribution, their temporal patterns are correctly identified as null. This is due to the incorporation of population-level variation in the NAWM. But this approach does not consider variation across other brain areas. To address this, we consider analyzing the normal variation across brain using control subjects in the next definition.

### 4.2.2 $H_{0,b}$ PPMS as Control

In order to detect differences in temporal patterns of abnormally enhancing MS lesions, one may also consider comparing the whole brain to controls. The ideal option would be to study DCE-MRI scans in healthy subjects. But the injection of a contrast agent is not completely risk-free and thus unethical to healthy subjects. One way around the problem is to identify subjects who were scanned, but whose DCE-MRI would be expected to be closest to that of a healthy brain.

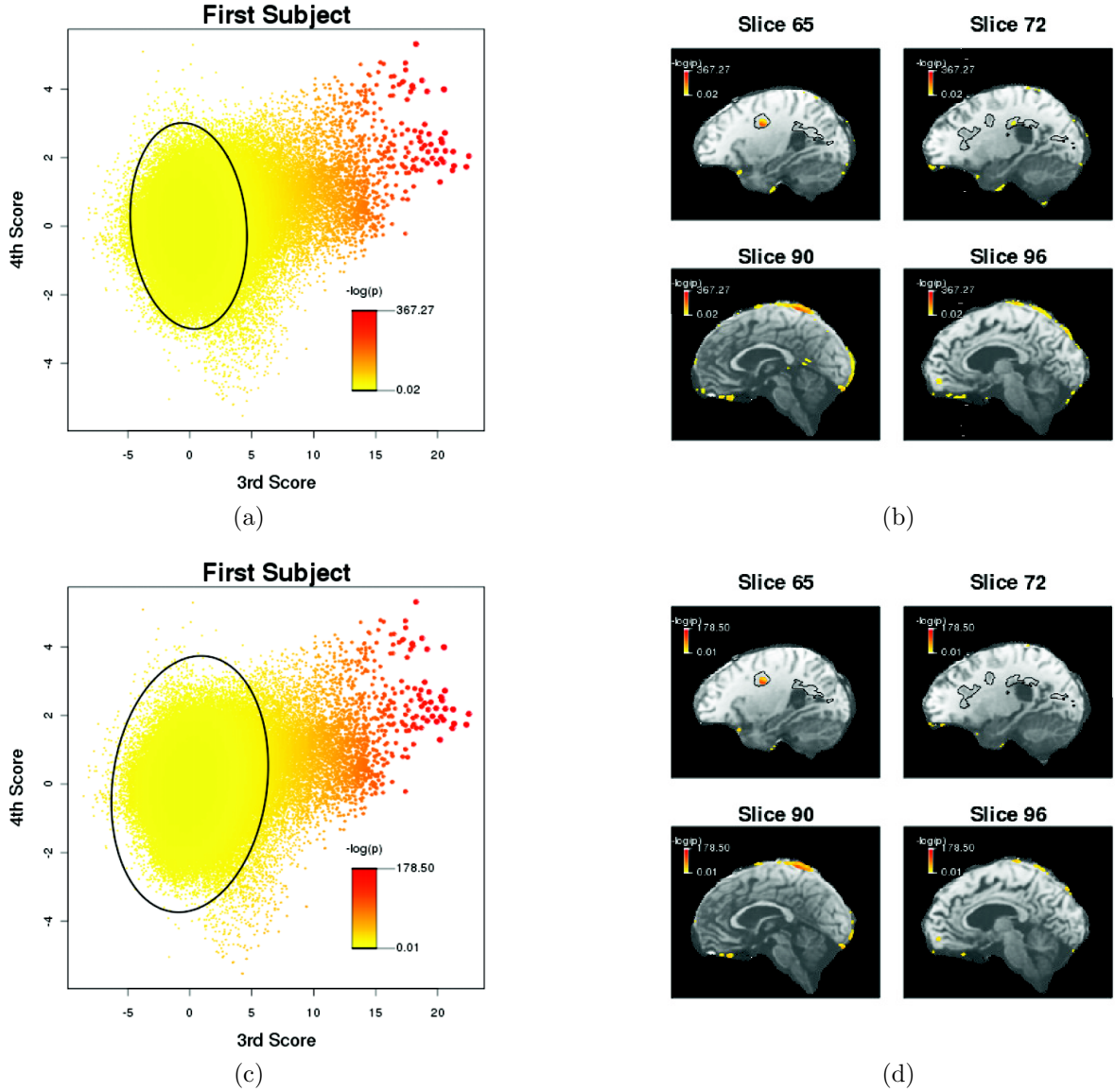

**Figure 9:** *Results of testing the first subject data on the population level null $H_{0,a}$ (a-b) and $H_{0,b}$ (c-d). $H_{0,b}$ provides a more conservative estimation of null distribution than $H_{0,a}$.*

As described in the introduction, lesion enhancement on DCE-MRI is seen primarily in RRMS patients, but seldom in PPMS subjects (Capra *et al.*, 1992). Thus, subjects with PPMS provide an alternative population that may be used as controls. Figure 10 compares the 3rd and 4th PC scores for all of the PPMS subjects (left) in our dataset and four randomly chosen subjects with RRMS. The red dots represent voxels in the lesion masks. Note that the PC scores for PPMS are concentrated around the origin while dots for RRMS lesions tend to exhibit arms shooting towards the top-right corner, indicating high PC 3 and PC 4 scores corresponding to enhancement. There are, however, some RRMS visits (the bottom-left plot in the right panel) without such arms, indicating subjects that are free of enhancing lesions at the time.

We formulate the null hypothesis with respect to PPMS patients as

$H_{0,b}$: the voxel-specific time series dynamics is the same with with those of brain voxels of PPMS subjects

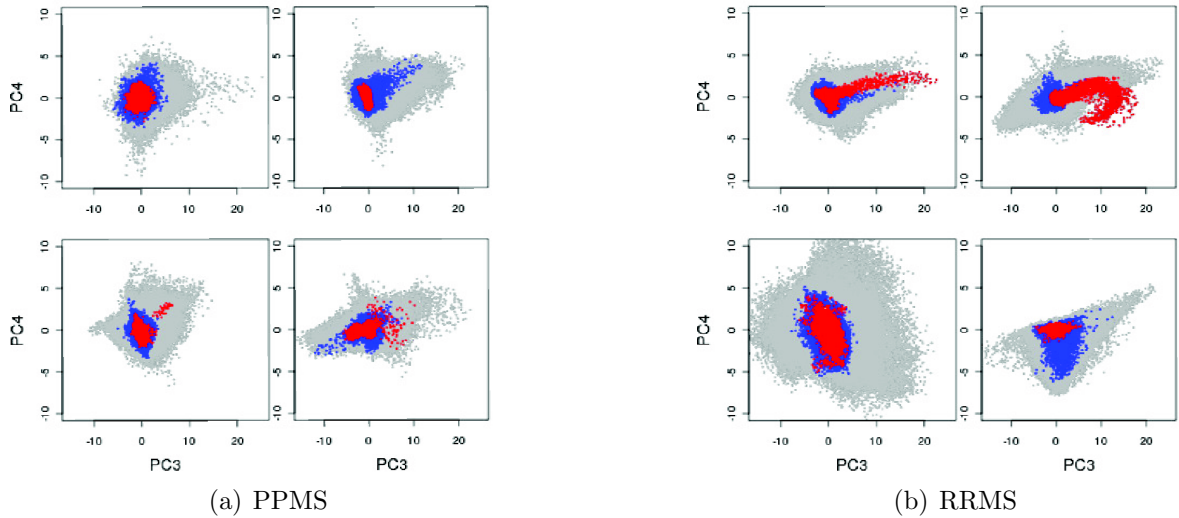

**Figure 10:** *Scatterplots of PC 3 vs. PC 4. The left sub-figure shows four PPMS subjects and the right one shows four RRMS subjects. The red points correspond to voxels in the lesion masks, the blue points are in the NAWM mask and the grey dots represent other voxels in the brain. Note that the PC scores for PPMS are concentrated around origin while those for RRMS subjects deviate farther from the origin. Typically, voxels in the lesion mask of the RRMS data have arms shooting towards the upright corner, indicating high PC3 and PC4 scores in enhancing lesions. However, there are also some RRMS subjects (such as that depicted in bottom left of (b)) without enhancing lesions which thus lack these arms.*

A closer look at $H_{0,b}$ indicates that voxels from the PPMS subjects are implicitly assumed to exhibit all the normal temporal behaviors except those corresponding to lesion enhancement. We also assume that voxels from RRMS patients follow a mixture distribution of these normal behaviors and enhancement behaviors. We obtain the stable density estimation of null distribution by fitting a normal distribution $N(\widehat{\mu}^{\text{PPMS}}, \widehat{\Sigma}^{\text{PPMS}})$ to PC scores in $\mathcal{A}_b = \{[\xi_{j,3}(\cdot), \xi_{j,4}(\cdot)], j \text{ is a PPMS patient}\}$. As before, we can conduct hypothesis testing on any data acquired using this protocol.

As the bottom row of Figure 9 shows, this approach seems to be quite conservative compared to the previous methods. When applied to subject 1, fewer voxels (only about 0.47%) of the brain are selected as enhancing. Among them 3.5% are in the lesion mask and 0.34% are in the NAWM mask. Voxels in lesion 2 are no longer classified as enhancing. $H_{0,b}$ incorporates subtle changes in intensity time series associated with the contrast injection that happens outside the lesions.

Although the results are promising, there are several concerns with this approach. First, it is based on the assumption that other than the enhancement in lesions, PPMS and RRMS patients are similar with respect to $T_1$-weighted MRI. This assumption is partially confirmed by comparing the distribution of 3$^{\text{rd}}$ and 4$^{\text{th}}$ PC scores for voxels outside the lesion masks. As displayed by the blue and gray dots in Figure 10, despite the between-sample variation, the bivariate PC scores have a similar shape for all samples. Second, only a small number of PPMS images are available in our study but larger future studies of DCE-MRI in patients with PPMS will be useful for refining the definition of non-enhancement through the null hypothesis $H_{0,\text{b}}$.

With all these candidate null hypotheses, when a new dataset is ready for testing, we could either select one of them according to the reliability of segmentation information, or we can test the data under all the null hypotheses and combine the results. We do not provide a strict criterion on how to draw final conclusion from all the results of the null hypothesis, as this is not the main aim of our paper. The goal is to make simple yes-or-no decisions while understanding what "yes" means and to understand the scientific problem from several new angles.

## 5 Conclusions

We have introduced the problem of identifying lesion voxels that enhance in DCE-MRI of MS patients. This problem is very complex because of the lack of a gold standard or precise definition of null hypothesis. This eminently qualitative hypothesis is then slowly transformed into a quantitative hypothesis using a close inter-play between data, statistical testing, and hard to quantify biological anatomy priors.

We proposed five ways of defining the null hypothesis for "non-enhancement" and estimating the corresponding null distributions. All five definitions aim at searching for data characteristics that comprehensively approximate normal behavior. While some are visit-specific, others use the entire sample. Some null distribution estimation procedures use segmentation information. When applied to data from subject 1, all of the above null hypotheses identify voxels in the enhancing lesions and confirm the existence of an additional (previously unknown) active lesion in the frontal lobe that was not detected by Lesion-TOADS. However, differences in specificity across the methods result in labeling of fewer or more NAWM voxels and voxels outside the WM mask as enhancing. We have found that many false positives are in the meninges and interstitial spaces and incorporating this anatomical information improves results.

We note that defining a null hypothesis is fundamentally different from the problem of multiple testing where a null hypothesis exists. To account for multiple testing we used Bonferroni correction instead of FDR (Benjamini & Yekutieli, 2001) because there is a one-to-one relationship between FDR and FWER $\alpha$ level. Neither of them solves the fundamental problem that we we do not know what the null hypothesis actually is. Instead we proposed a spectrum of null hypotheses that can be explored, criticized, and refined. We have labeled this paradigm the "soft null hypothesis", to emphasize that data can and should be used to define and refine null hypotheses.

## SUPPLEMENTAL MATERIALS

**Title:** Supplementary Materials for Soft Null Hypothesis Testing in Lesion Detection (PDF)

**A. Comparison of the Smoothing Methods on PLPCs** Results comparing the simple linear interpolation described in Section 2 of the manuscript and the spline smoothing method for raw intensity curves for each voxel. We conduct principal component analysis on intensity curves that are smoothed with the two methods. The resulting population level principal components (PLPCs) are shown to be similar.

**B. Results for Clustering Analysis** Detailed results of clustering analysis based on the $3^{\text{rd}}$ and $4^{\text{th}}$ PC scores. We cluster the voxels by fitting a mixture normal distribution with $3 \sim 6$ components and compare the results. We also take the fitted distributions as null distribution and obtain the testing results with respect to the null.

**C. Soft Null Hypothesis Test for Subject 2** Testing results for a second subject under the soft null procedure.

# References

Benjamini, Y., & Yekutieli, D. 2001. The control of the false discovery rate in multiple testing under dependency. *The Annals of Statistics*, **29**, 1165–1188.

Capra, R., Marciano, N., Vignolo, L.A., Chiesa, A., & Gasparotti, R. 1992. Gadolinium-pentetic acid magnetic resonance imaging in patients with relapsing remitting multiple sclerosis. *Archives of Neurology, Chicago*, **49**, 687–689.

Carass, A. 2007. A joint registration and segmentation approach to skull stripping. *Proceedings of the 2007 IEEE International Symposium on Biomedical Imaging: From Nano to Macro. Washington, DC,USA. April 12-16, 2007*, 656–659.

Davidian, M., & Giltinan, D.M. 1995. *Nonlinear models for repeated measurement data*. Chapman & Hall/CRC0412983419.

Efron, B. 2004. Large-scale simultaneous hypothesis testing: the choice of a null hypothesis. *Journal of the American Statistical Association*, **99**, 96–104.

Efron, B. 2007. Correlation and large-scale simultaneous significance testing. *Journal of the American Statistical Association*, **102**, 93–103.

Fox, R.J., Bethoux, F., Goldman, M.D., & Cohen, J.A. 2006. Multiple sclerosis: advances in understanding, diagnosing, and treating the underlying disease. *Cleveland Clinic Journal of Medicine*, **73**, 91–102.

Gaitán, M.I., Shea, C.D., Evangelou, I.E., Stone, R.D., Fenton, K.M., Bielekova, B., Massacesi, L., & Reich, D.S. 2011. Evolution of the blood–brain barrier in newly forming multiple sclerosis lesions. *Annals of Neurology*.

Li, X., Rooney, W. D., & Springer, C. S. 2005. A unified magnetic resonance imaging pharmacokinetic theory: intravascular and extracellular contrast reagents. *Magnetic Resonance in Medicine*, **54**, 1351–1359.

Lublin, F.D., & Reingold, S.C. 1996. Defining the clinical course of multiple sclerosis. *Neurology*, **46**, 907–911.

Shiee, N., Bazin, P.L., Reich, D., & Pham, D.L. 2010. A topology preserving approach to the segmentation of brain images with multiple sclerosis lesions. *Neuroimage*, **49**, 1524–1535.

Shinohara, R.T., Crainiceanu, C.M., Gaitán, M. I., Caffo, B.S., & Reich, D. 2011. Population-wide principal component-based quantification of blood-brain-barrier dynamics in multiple sclerosis. *Neuroimage*, **57**, 1430–46.

Tofts, P.S., Brix, G., Buckley, D.L., Evelhoch, J.L., Henderson, E., Knopp, M.V., Larsson, H.B.W., Lee, T., Mayr, N.A., Parker, G.J.M., Port, R.E., Taylor, J., & Weisskoff, R.M. 1999. Estimating kinetic parameters from dynamic contrast-enhanced T(1)-weighted MRI of a diffusable tracer: standardized quantities and symbols. *Journal of Magnetic Resonance Imaging*, **10**, 223–232.

Yankeelov, T.E., Luci, J.J., Lepage, M., Li, R., Debusk, L., Lin, P.C., Price, R.R., & Gore, J.C. 2005. Quantitative pharmacokinetic analysis of DCE-MRI data without an arterial input function: a reference region model. *Magnetic resonance imaging*, **23**, 519–29.

# Supplementary Materials for Soft Null Hypothesis Testing in Lesion Detection

June 24, 2013

## A. Comparison of the Smoothing Methods on PLPCs

In our analysis, we adopt the simple linear interpolation method, i.e., connecting the intensity curve by a line between two observed points. We use the linear interpolation based on two assumptions: 1) the intensities change continuously over time after injection; 2) the timing is dense enough for each dataset to capture the trend, i.e., there is no unobserved peak or valley between two time points. To assess the reliability of using linear interpolation, we experimented other spline interpolation method such as natural splines and obtained a set of new PCs (shown in Figure 1). Although this results in smoother appearing components than the linearly interpolated PCs, they capture similar patterns of the intensities. Moreover, the natural spline interpolation is computationally more expensive (smoothing procedure takes 0.1s more per voxel). Considering the large number of voxels within one brain, such a smoothing approach may be inefficient and unnecessary. Meanwhile, the disadvantage of using quadratic or cubic basis functions is that they often have poor performance at the two ends of the curve as shown in the Figure 1. Indeed, it is scientifically known that the left end of the curves should stay flat because the contrast agent is absent.

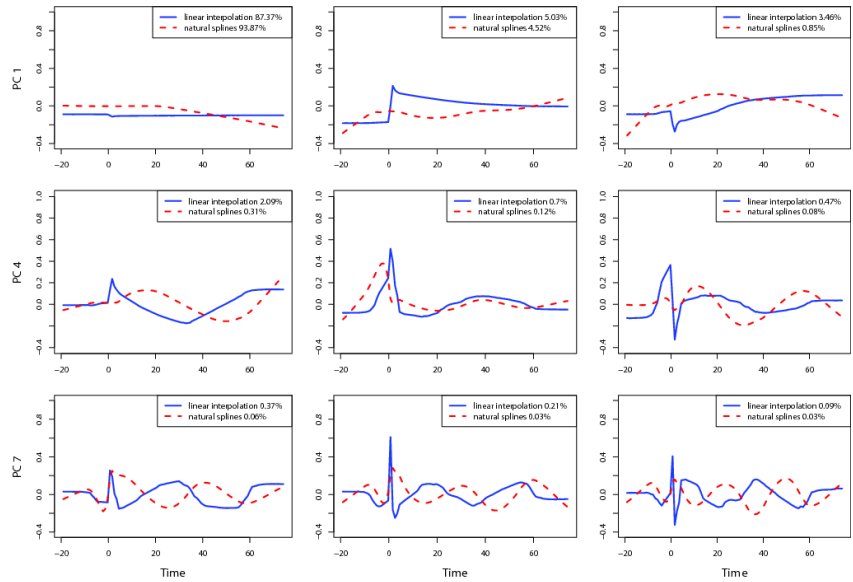

Figure 1: The first nine principal components obtained from data that are linearly interpolated (solid blue lines) and from splines (dashed red lines). The percentages in the legend are the proportions of variance explained by each component.

## B. Results for Clustering Analysis

We conducted clustering analysis by fitting a k-mixture normal distribution to the bivariate $3^{\text{rd}}$ and $4^{\text{th}}$ PC scores. The results under different number of components k (k = 3, 4, 5, 6) are demonstrated in Figure 2. In each panel, colors represent the estimated clusters. The black ellipsoids are the 95% percentile contours Given the huge number of points to be fitted on the scatter plot, a mixture normal distribution with 6 components is preferred with the lowest log likelihood. However, note that from a 5-mixture normal distribution to a 6-mixture normal distribution, the one additional component is estimated to be concentrated around the origin. It does not help with identifying the enhancing lesion voxels that are known to be scattered on the boundary of the cloud. In fact, the 4-mixture normal distribution has one component that push the furthest to the boundary. We map the clustered voxels back to the brain and obtain the green cluster in 4-mixture normal as shown in Figure 3. The results contain more false positive than what we obtain using the population-level null hypotheses proposed in the paper. The reason that clustering is not able to separate the enhancing voxels as an individual cluster is that: 1) there is a continuous change in intensity for voxels going from no enhancement to strong enhancement. Although the enhancing lesion voxels in general demonstrate strong intensity signal and therefore high loadings on the $3^{rd}$ and $4^{th}$ principal components, they do not separate away from other voxels and form an individual cluster. 2) Lesion voxels only account for a small portion of all voxels in the brain. Therefore those voxels do not have high influence when estimating the parameters for mixture distributions. This is why we take a hypothesis testing approach in the paper instead of clustering analysis.

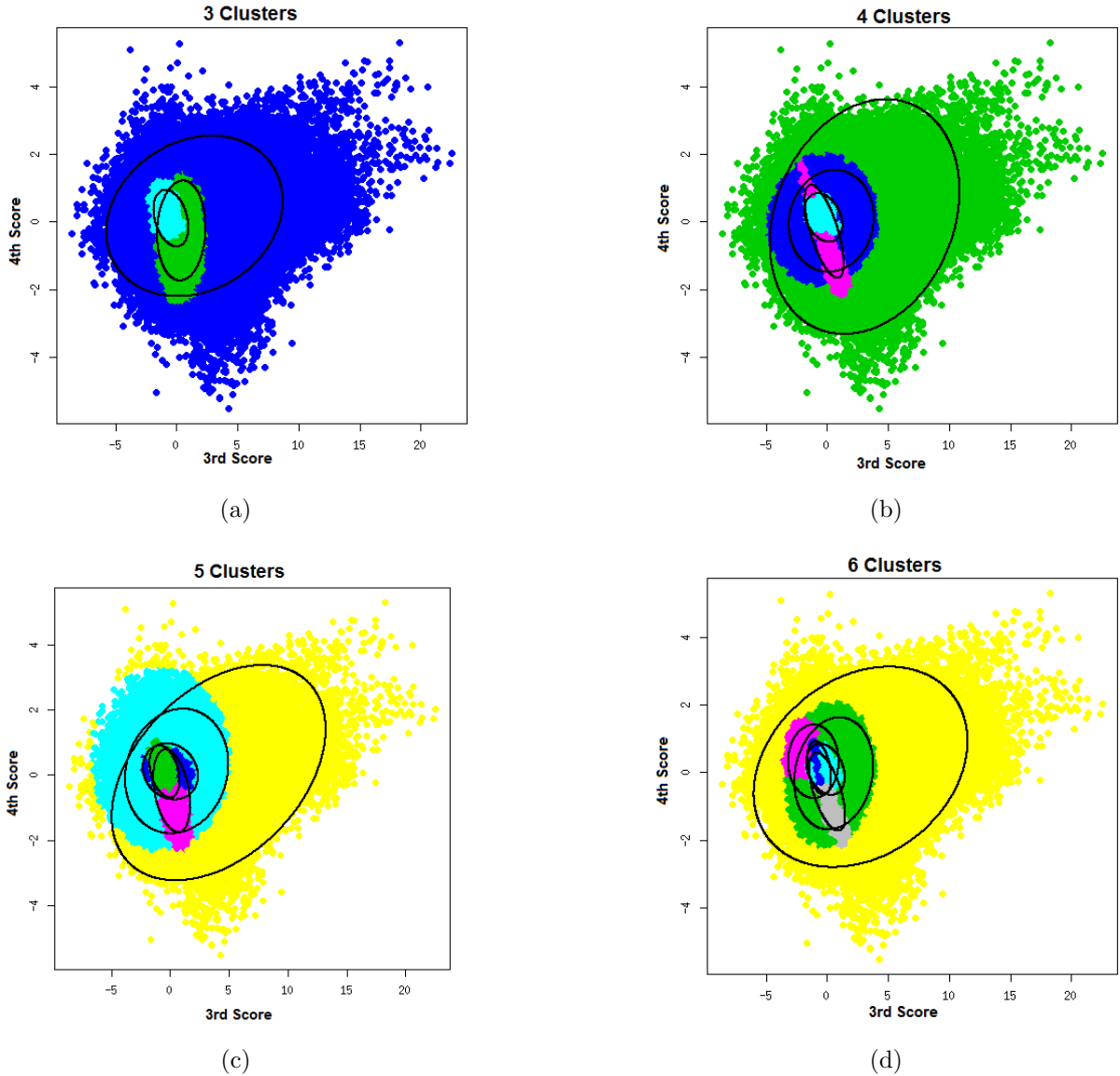

Figure 2: Clustering results when fitting three to six mixture normal components. Colors representing clusters. The black ellipsoids are the 95% percentile contours for each normal component.

Alternatively, to examine whether the soft null procedure is necessary, we take the fitted mixture distributions as null distribution for nonenhancement, given that enhancing lesion voxels that we aim to detect has low leverage in determining the mixture distribution. The p-values of voxel $v$

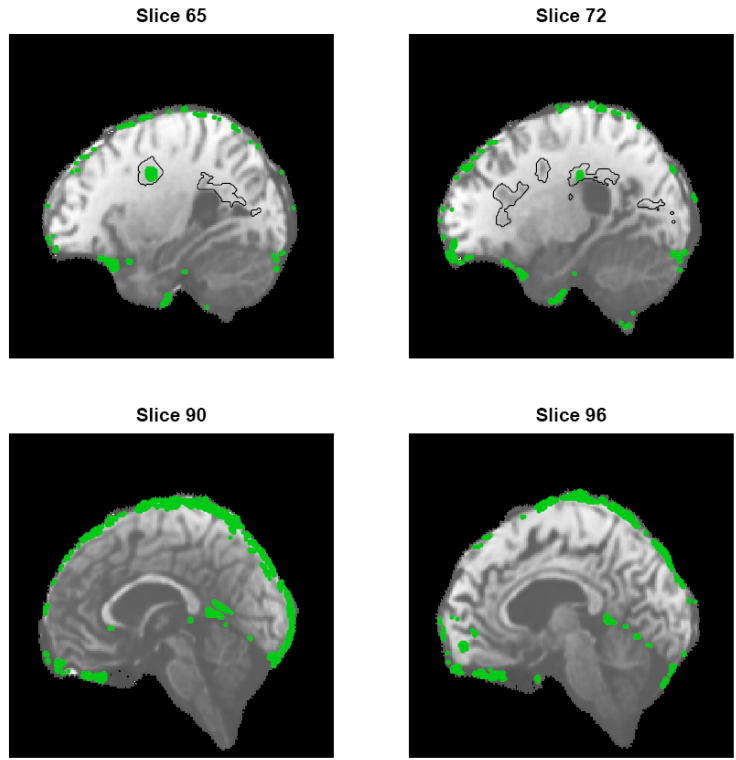

Figure 3: The green dots on the imaging slices indicate voxels that are clustered as enhancing lesion voxels by fitting a 4-mixture normal distribution.

against a k-mixture distribution $\sum_{j=1}^{k} w_j f_j$ are determined as

$$p_v = \sum_{j=1}^{k} w_j p_{jv} \qquad (1)$$

where $w_j$ is the weight of the $j^{\text{th}}$ normal and $p_{jv}$ is the p-value of voxel $v$ against the $j^{\text{th}}$ normal. The p-values are on a larger magnitude than fitting only one normal distribution, which is expected. However, under different threshold levels, the distinguish of true positive and false positive is not obvious. Under the threshold of 0.05 (Figure 4), 3.12% (46813) of the voxels in the brain are selected as enhancing lesion voxels. However only 1% of them overlap with the lesion masks provided by Lesion-TOADS algorithm. When we reduce the threshold to be 0.03, fewer noise voxels as well as lesion voxels are detected below the threshold.

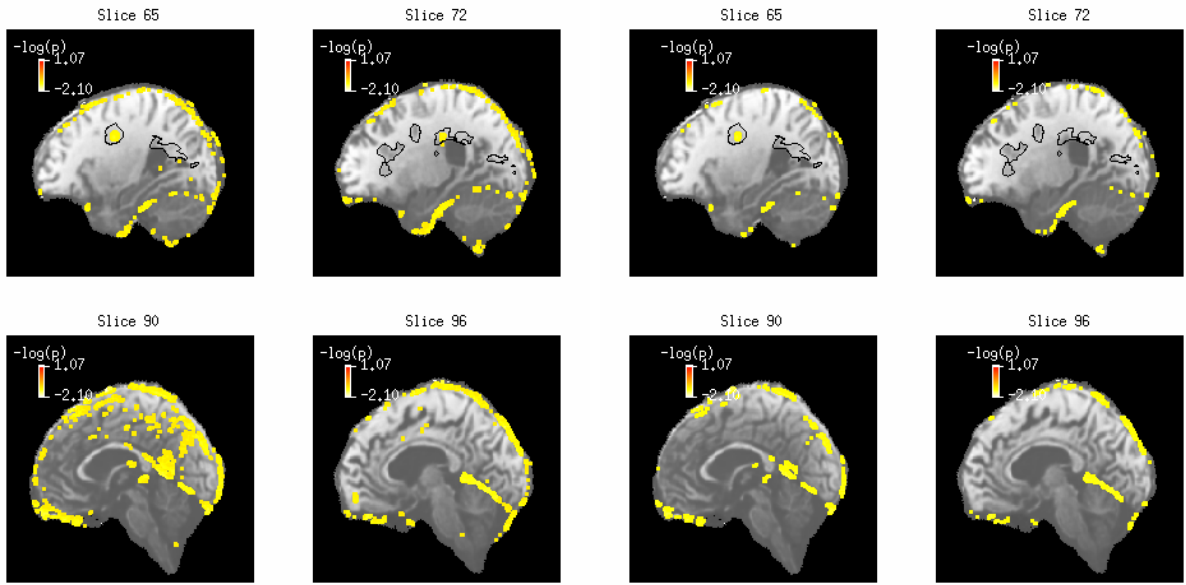

Figure 4: The left panel show the voxels that are tested to be significantly enhancing when the null distribution is the 4-mixture normal fitted based on the whole dataset. The colored dots are voxels with p-values under 0.05. On the left panel are voxels with p-values under 0.03. No voxel has p-value under 0.01.

## C. Soft Null Hypothesis Test for Subject 2

In this section, we provide analogous plots as Figure 8,9,10 in the manuscript for testing results on a different subject. This subject has an enhancing lesion with open ring shaped. The original MR images are shown as the middle panel of Figure 1 in the manuscript. Similar to what we observed from subject 1 that is discussed in the manuscript, applying the five null hypotheses dynamically help us to understand the various enhancement patterns across different tissue types in the brain of subject 2. Moreover, comparing the plots for subject 1 and subject 2, we may obtain certain understanding about the subject-to-subject variation. Also note that the nodular lesion detected on slice 109 actually locates right below the lesion mask (black contour), indicating that our method may be a more reliable approach than the visual inspection and manual segmentation.

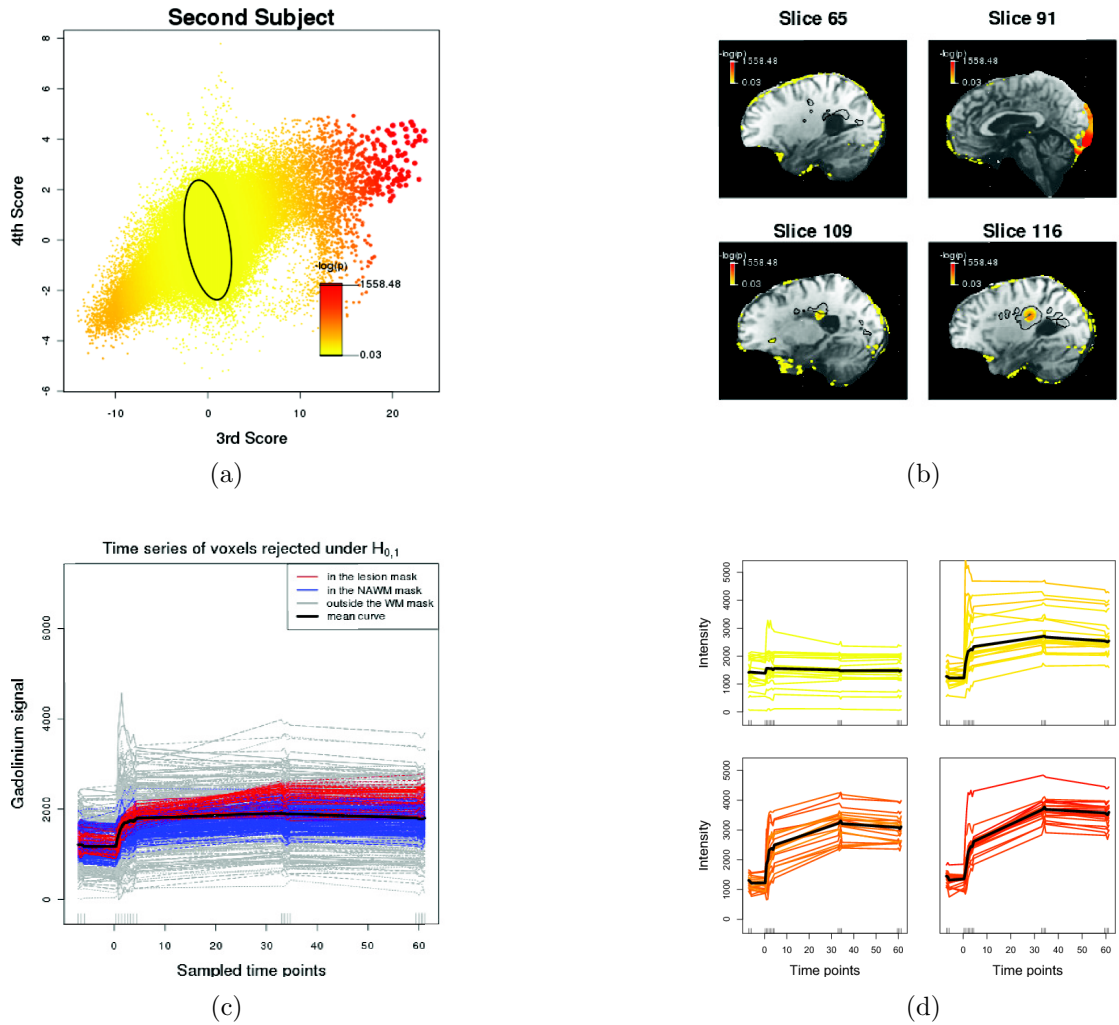

Figure 5: Analysis results of data for subject 2 under $H_{0,1}$. (a) The $p$-values of each voxel is represented in color. The size of red dots are amplified for clarity in illustration. The black ellipsoid is the 95% confidence contour under the Bonferroni correction of the total number of voxels that are tested; (b) shows the location of rejected voxels in 4 different slices of the brain image. Slice 65 shows older lesions that do not enhance. Slice 91 shows some blood vessels located at the back of the brain that have extreme large PC 3 and PC 4 scores. On slice 109, some small lesions are selected. Slice 116 clearly shows the ring shaped lesion with highly significant $p$-values; (c) are the time series of a subset of voxels rejected under $H_{0,1}$. Those colored red are in the lesion mask and the blue curves corresponds to voxels in the NAWM mask. The grey lines represent voxels outside WM mask; (d) we stratify the voxels by their $p$-values into 4 groups and draw 20 sample series in each group as well as the mean curve (black). The upper-left graph shows for voxels that are not rejected. Note that voxels with smaller $p$-values show a larger slope.

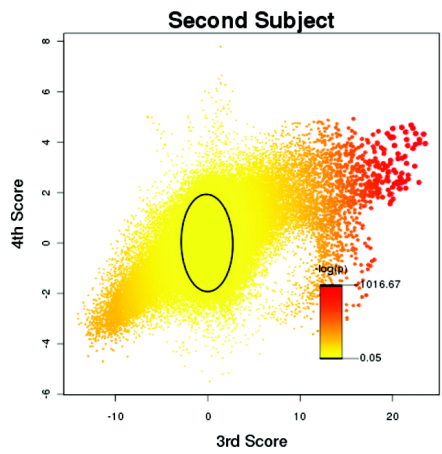

(a)

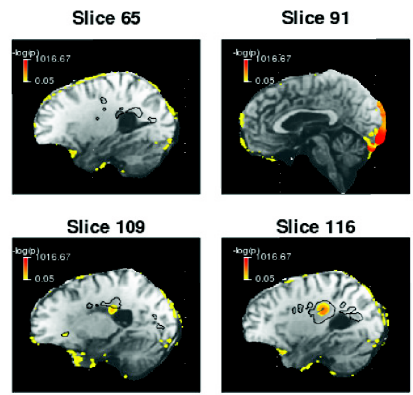

(b)

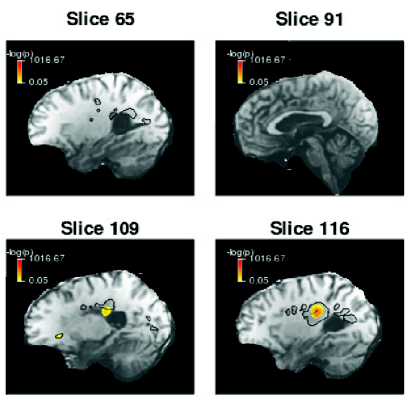

(c)

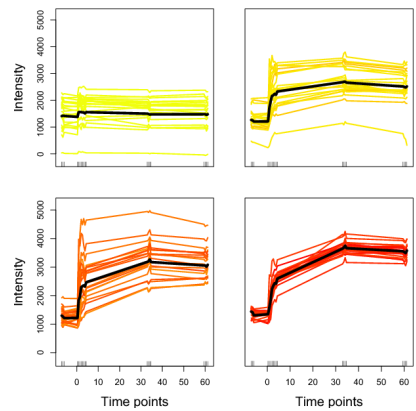

(d)

Figure 6: Testing results of subject 2 under $H_{0,2}$. (c) only shows voxels that are show significant enhancement but lie outside the NAWM mask. (a),(b) and (d) are the same plots shown in Figure 5.

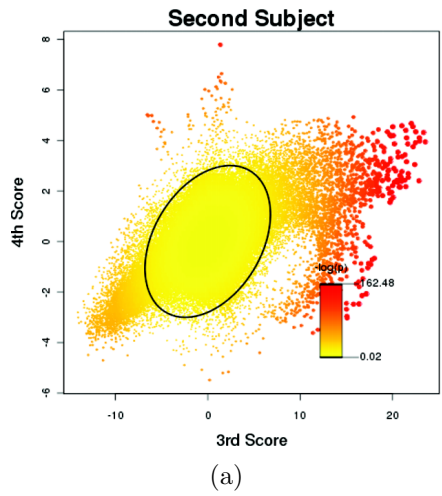

(a)

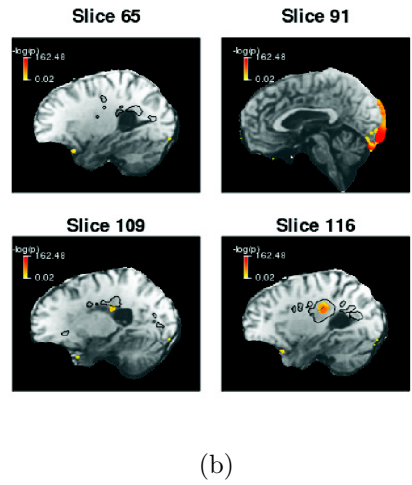

(b)

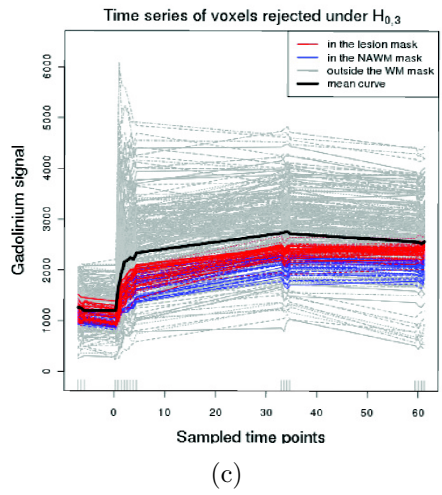

(c)

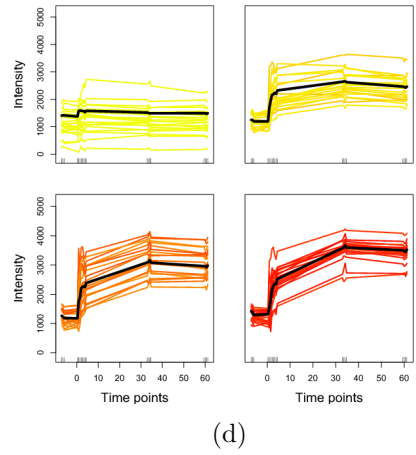

(d)

Figure 7: Testing results of subject 2 under $H_{0,3}$. Same plots as Figure 5.

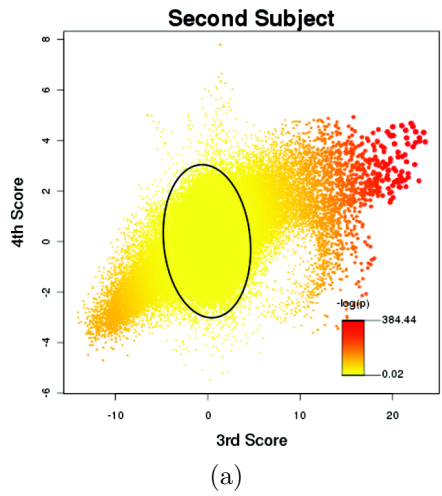

(a)

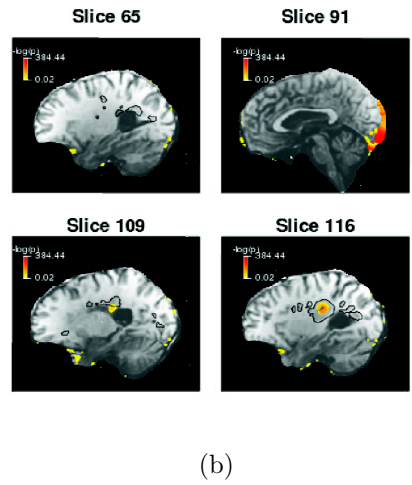

(b)

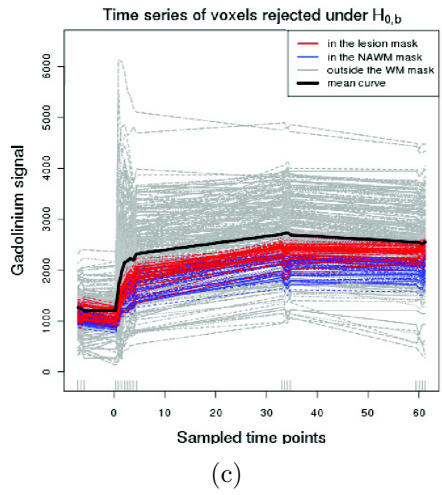

(c)

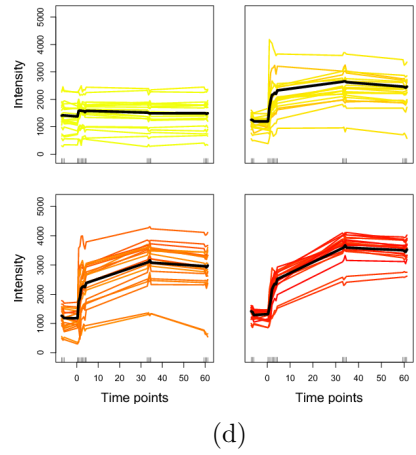

(d)

Figure 8: Testing results of subject 2 under $H_{0,a}$. Same plots as Figure 5.

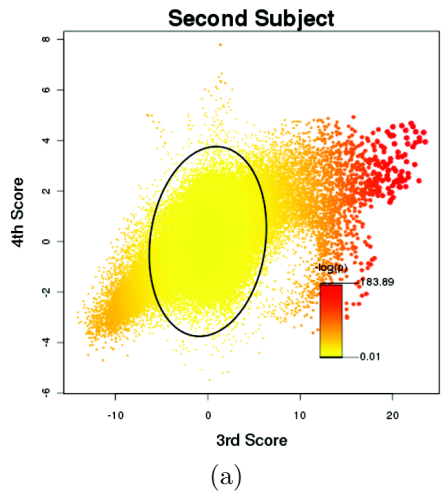

(a)

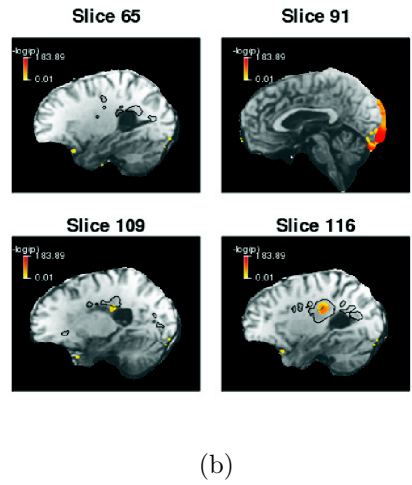

(b)

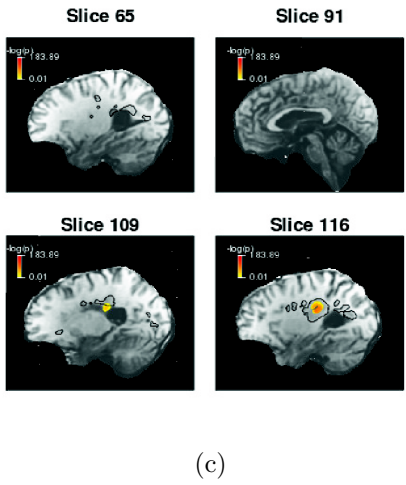

(c)

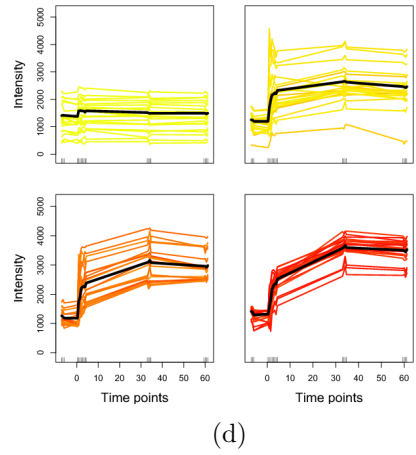

(d)

Figure 9: Testing results of subject 2 under $H_{0,b}$. Same plots as Figure 6.